\documentclass[prb,twocolumn,showpacs,amsmath,amssymb]{revtex4}

\usepackage{graphicx}
\usepackage{dcolumn}
\usepackage{bm}

\begin{document}

\preprint{APS/123-QED}

\title{Magnetic field-induced quantum critical point in YbPtIn and YbPt$_{0.98}$In single crystals}

\author{E. Morosan,$^{1,2}$ S. L. Bud'ko,$^{1,2}$ Y. A. Mozharivskyj $^{2,3}$ and P. C. Canfield $^{1,2}$}
\affiliation{$^1$Department of Physics and Astronomy, Iowa State
University, $^2$Ames Laboratory, $^3$Department of Chemistry, Iowa
State University, Ames, IA 50011, USA}

\date{\today}

\begin{abstract}
Detailed anisotropic (H$\parallel$ab and H$\parallel$c) resistivity
and specific heat measurements were performed on online-grown YbPtIn
and solution-grown YbPt$_{0.98}$In single crystals for temperatures
down to 0.4 K, and fields up to 140 kG; H$\parallel$ab Hall
resistivity was also measured on the YbPt$_{0.98}$In system for the
same temperature and field ranges. All these measurements indicate
that the small change in stoichiometry between the two compounds
drastically affects their ordering temperatures (T$_{ord}\approx3.4$
K in YbPtIn, and $\sim2.2$ K in YbPt$_{0.98}$In). Furthermore, a
field-induced quantum critical point is apparent in each of these
heavy fermion systems, with the corresponding critical field values
of YbPt$_{0.98}$In (H$^{ab}_c$ around 35-45 kG and
H$^{c}_c\approx120$ kG) also reduced compared to the analogous
values for YbPtIn (H$^{ab}_c\approx60$ kG and H$^{c}_c>140$ kG).

\end{abstract}


\maketitle

\section{Introduction}
In recent years, stoichiometric Yb-based heavy fermion compounds
have raised a lot of interest, particularly due to the limited
number of such systems known to date. YbPtIn is one of the few
examples of such heavy fermion systems\cite{tro09,kac10,yos11}, as
indicated by measurements performed on single crystal samples
extracted from on-line melts of the polycrystalline material. Based
on the existing data\cite{tro09,kac10}, YbPtIn has a relatively low
magnetic ordering temperature (T$_{ord}~\sim~3$ K) and an enhanced
electronic specific heat coefficient ($\gamma~>~400$ mJ$/$mol
K$^2$), whereas the magnetic entropy at T$_{ord}$ amounts to only
about 60$\%$ of the $R~\ln2$ value expected for a doublet ground
state. In light of these observations, this system appears to be
qualitatively similar to the other recently studied Yb-based heavy
fermion antiferromagnets,
YbRh$_2$Si$_2$\cite{tro05,geg06,ish07,pas08,pas09}, and
YbAgGe\cite{bey27,kat28,mor05,bud11,bud12}, the latter compound
being isostructural to YbPtIn. In YbRh$_2$Si$_2$, magnetic order
occurring at very low temperature (below 70 mK)\cite{tro05} was
associated with a low entropy release (around 0.01
$*~R~\ln2$)\cite{geg06}. For the YbAgGe\cite{mor05,bud11} compound,
the temperature associated with the magnetically ordered state,
whereas still fairly low ($\sim~1.0$ K), is enhanced compared to
that of YbRh$_2$Si$_2$; also, the magnetic entropy at the ordering
temperature is larger than in YbRh$_2$Si$_2$, but still less than
10$\%$ of $R~\ln2$. Thus these two compounds can be regarded as
systems with small moment ordering. Having an even higher ordering
temperature and magnetic entropy at T$_{ord}$, YbPtIn seemed a good
candidate to further study a progression from small moment to
reduced moment ordering, in stoichiometric Yb-based heavy fermion
compounds, with field-induced quantum critical point.

In contrast to a classical phase transition at finite temperatures,
driven by temperature as a control parameter with thermal
fluctuations, a quantum phase transition is driven by a control
parameter C other than temperature (\textit{e.g.}, C = pressure,
doping or magnetic field) at T = 0, with quantum mechanical
fluctuations. Such a control parameter tunes the system from a
magnetically ordered state towards a state without magnetic order,
at zero temperature, crossing a quantum critical point. Due to the
hybridization of the 4f electrons and the conduction electrons in
heavy fermion (HF) systems, which can be modified by any one of the
aforementioned C control parameters, the (HF) compounds are very
suitable to study quantum critical behavior. Moreover, close to the
critical value C$_{crit}$ which drives the ordering temperature
close to zero, pronounced deviations from the Fermi liquid-like FL
behavior can occur. This has been observed in a large number of HF
systems where C = doping or pressure, and only a few doped systems
have been field-tuned through a QCP\cite{ste2001}. To date,
YbRh$_2$Si$_2$\cite{tro05,geg06,ish07,pas09} and
YbAgGe\cite{kat28,bud11,bud12} are the only stoichiometric Yb-based
HF compounds in which a field-induced quantum critical point QCP has
been observed. The heavy fermion character of the YbPtIn system has
already been reported\cite{tro09,kac10}; in this paper we will
discuss the existence of a field-induced QCP in YbPtIn which renders
it very similar to the other two stoichiometric, Yb-based heavy
fermions.

Recently we presented anisotropic low-field susceptibility
measurements, as well as specific heat data in zero applied field,
on solution-grown single crystals of the RPtIn series, including
YbPtIn\cite{mor10}. No features indicative of magnetic order could
be identified in our magnetization measurements on single crystals
of the R = Yb compound, down to T $\sim$ 2 K. However, a
well-defined peak at $T~=~2.1$ K in the zero-field specific heat
data suggests that this compound orders magnetically below 2.1 K,
just above the low temperature limit of our magnetization
measurements. As we will present in this paper, detailed transport
and thermodynamic measurements down to 0.4 K not only confirm the
magnetic ordering in this compound below 2.1 K, but also suggest
that another phase transition might exist in this system around 1 K.

However, our measurements were in part inconsistent with previous
reports on YbPtIn single crystals: whereas Kaczorowski \textit{et
al.}\cite{kac10} have also presented magnetization data showing no
features associated with magnetic ordering above 1.7 K, in their
specific heat measurements three anomalies can be observed (at 3.1,
2.3, and 1.2 K), all at slightly different temperatures than in our
data. Furthermore, the low-field susceptibility data presented by
Trovarelli \textit{et al.}\cite{tro09} are consistent with an
ordering temperature around 3.4 K, also confirmed by their
resistivity and specific heat measurements.

In order to address these apparent discrepancies between our data on
solution grown single crystals and the two previous reports on
on-line grown single crystals\cite{tro09,kac10}, we reproduced the
growth as described by Trovarelli \textit{et al.}\cite{tro09}. The
anisotropic specific heat and transport measurements on our single
crystals extracted from the melt confirmed the existence of magnetic
phase transitions at 3.4 K and 1.4 K, as observed previously.

In this paper we will try to examine the differences between
solution and on-line grown single crystals, given that a small
Pt-deficiency occurred in the former types of crystals, leading to a
stoichiometry closest to YbPt$_{0.98}$In, and no disorder could be
detected in the on-line grown crystals. Also, given that heavy
fermion compounds with small moment ordering are likely to be driven
to a quantum critical point QCP by disorder or applied field, we
will study the evolution of both YbPt$_{0.98}$In and YbPtIn towards
a field-induced QCP. Detailed anisotropic measurements of specific
heat and resistivity on the two types of crystals, for fields up to
140 kG and temperatures down to 0.4 K, will be used for the
comparison between the two types of crystals. Additional Hall effect
measurements were performed on the solution grown single crystals
for the same temperature and field ranges, allowing us to further
explore the effects of the QCP in this compound.

After briefly describing the methods used for synthesizing the two
types of YbPtIn single crystals, and the measurement techniques used
for the sample characterization, we will present the results of
single crystal x-ray diffraction. Next we will compare the low field
data for the two types of compounds, pointing out the similarities,
as well as the significant differences in their physical properties.
Next we will present the higher field data for both types of
crystals, for $H\parallel~ab$ and $H\parallel~c$. This will allow us
to follow, in both systems, the progression from reduced magnetic
moment order to the quantum critical point QCP, as driven by the
application of increasing magnetic field. Also, we will extend the
zero-field comparison between the two YbPtIn systems to their field-
dependent properties, and will try to identify common features as
well as possible effects of the site deficiency.

\section{Experimental methods}

YbPtIn is one of the RPtIn compounds, reported to crystallize in
ZrNiAl hexagonal structure\cite{kac10,mor10}, space group
P$\overline{6}2m$. Single crystals of this compound were obtained by
two methods: from high-temperature ternary solution, as described in
Ref. [15], and from stoichiometric on-line melt\cite{tro09}. The
initial concentration used for the solution growth was
Yb$_{0.4}$Pt$_{0.1}$In$_{0.5}$, and the ternary solution, sealed in
a 3-cap Ta crucible\cite{can17} under partial argon pressure, was
slow-cooled from 1200$^0$ C to 1000$^0$ C over approximately 100
hours. Subsequently, the excess liquid solution was decanted, and
the resulting hexagonal rods were briefly etched in HCl to remove
residual flux from the surface. For the on-line melt, 1:1:1 atomic
ratios of Yb, Pt and In were sealed in a Ta crucible under partial
argon atmosphere. After briefly heating the sealed crucible up to
approximately 1650$^0$ C in an induction furnace, the crucible was
sealed in a quartz ampoulle and annealed at 700$^0$ C for 120 hours.
Small rods were subsequently extracted from the resulting
crystalline conglomerate.

Anisotropic magnetization measurements as a function of temperature
and applied field M(H,T) were performed in a Quantum Design MPMS
SQUID magnetometer (T $=~1.8~-~350$ K, H$_{max}~=~55$ kG).
Temperature- and field-dependent specific heat C$_P$(H,T),
resistivity $\rho$(H,T) and Hall resistivity $\rho_H$(H,T)
measurements were taken in a Quantum Design PPMS-14 instrument with
He-3 option, for temperatures down to 0.4 K, and applied magnetic
fields between 0 and 140 kG. For the anisotropic (\textit{i.e.,}
$H\parallel ab$ or $H\parallel c$) specific heat measurements a
relaxation technique with fitting of the whole calorimeter (sample
with sample platform and puck) was used. The sample platform and
grease background, measured separately for all necessary (H,T)
values, was subtracted from the sample response. Also, the
non-magnetic contribution to the total specific heat was estimated
based on measurements on flux-grown LuPtIn, used as the non-magnetic
analogue of YbPtIn. The single crystals of LuPtIn were grown using
similar initial composition (R$_{0.05}$Pt$_{0.05}$In$_{0.90}$) and
temperature profile (slow cooling from 1190$^0$C to 800$^0$C over
$\sim$100 hours) that yielded fully-occupied RPtIn single crystals
for the lighter members of the series\cite{mor10}. We therefore
anticipate that the LuPtIn crystals are also fully-occupied, and we
used this system as the non-magnetic analogue for our both Yb
compounds. It should be noted that we made the assumption that the
non-magnetic background is insensitive to the 2$\%$ Pt-deficiency in
the solution-grown compound. The specific heat for LuPtIn was
measured in the same temperature range as for YbPtIn and
YbPt$_{0.98}$In, for 0 and 140 kG applied field, and was found to be
virtually field-independent over the measured temperature range.

A standard AC four-probe resistance technique ($f~=~16$ Hz,
$I~=~1-0.3$ mA) was used for the field- and temperature-dependent
resistivity measurements. Given the rod-like geometry of the
samples, the current was always flowing along the crystallographic
c-axis, whereas the field was either parallel (for longitudinal
magnetoresistance), or perpendicular (for transverse
magnetoresistance) to the direction of current flow. The Hall
resistivity measurements were performed on a sample that was
polished down to a plate-like shape. This sample geometry restricted
the Hall resistivity measurements to the field applied within the
hexagonal crystallographic plane ($H~\parallel~ab$). The Hall
resistivity $\rho_H$(H,T) was measured for field perpendicular to
both the current and the Hall voltage directions. In order to
minimize the inherent (small) misalignment of the voltage contacts,
these measurements were taken for two opposite directions of the
applied field, H and -H, and the odd component, ($\rho_H$(H) -
$\rho_H$(-H))$/2$ was taken as the Hall resistivity.

For each of the two systems (YbPtIn and YbPt$_{0.98}$In), we will
compare the similar data for the the two orientations of the applied
field (\textit{i.e.}, for $H~\parallel~ab$ and $H~\parallel~c$), as
well as the resulting T - H phase diagrams. The criteria used for
determining the points on all presented phase diagrams will be
discussed in more detail in the appropriate sections, but in general
they were maxima in the C$_P$(H,T) data, and derivative maxima or
onset values for the transport data. For $H~\parallel~ab$, the
maximum temperature for which the $\rho(T^2)$ could be fit to a
straight line was the criterium used in determining the upper limit
of this Fermi liquid-like region. Additional Hall resistivity
measurements were performed for the field applied within the basal
plane, for the YbPt$_{0.98}$In (flux-grown) compound. Various
criteria were used for determining critical points similar to the
Hall line for the YbRh$_2$Si$_2$\cite{pas09} and YbAgGe\cite{bud12}
systems.

\section{Results}

\subsection{Crystal structure}

For both solution and on-line grown crystals, the crystal structure
was confirmed by powder x-ray diffraction, with no detectable
impurity peaks. However, additional single crystal x-ray
measurements were performed on the two types of YbPtIn compounds.
Crystals with dimensions $\sim2~\times~7~\times~11~\mu m^3$ and
$\sim2~\times~6~\times~13~\mu m^3$ were extracted from the flux and
on-line grown samples respectively. Room-temperature X-ray
diffraction data were collected on a STOE IPDSII image plate
diffractometer with Mo K$\alpha$ radiation, and were recorded by
taking $1^0$ scans in $\theta$ in the full reciprocal sphere. The
range of 2 $\theta$ extended from $6^0$ to $63^0$. Numerical
absorption corrections for both crystals were based on crystal face
indexing, followed by a crystal shape optimization. Structure
solution and refinement were done using the SHELXTL program. The
crystallographic and structural data are summarized in Tables
\ref{t1} - \ref{t2}.

\begin{table}[htb]
 \begin{center}
 \caption{Atomic coordinates and equivalent isotropic displacement parameters
($\AA^2$) for the flux-grown YbPtIn system. U(eq) is defined as one
third of the trace of the orthogonal U$^{ij}$ tensor. Space group
\textit{P $\overline{6}$ 2 m}, a $=~7.5568(8)~\AA$, c
$=~3.7658(3)~\AA$, R = 0.0273, R$_\omega$ = 0.0598.} \label{t1}
\begin{tabular}{|c|c|c|c|c|c|} \hline
$~$&$~$&$~$&$~$&$~$&$~$ \\
& Occupancy & x & y & z & U(eq)\\
$~$&$~$&$~$&$~$&$~$&$~$ \\
\hline $~~~$&$~~~$&$~~~$&$~~~$&$~~~$&$~~~$ \\
$~~~$Yb$~~~$ & $~~~$1.00$~~~$& 0.5940(1) & $~~~$0$~~~$ & $~~~$0$~~~$ & 0.0088(3) \\
$~$&$~$&$~$&$~$&$~$&$~$ \\
Pt(1) & 0.94(1) & 0 & 0 & 0 & 0.0099(5) \\
$~$&$~$&$~$&$~$&$~$&$~$ \\
Pt(2) & 1.00 & 1/3 & 2/3 & 1/2 & 0.0075(3) \\
$~$&$~$&$~$&$~$&$~$&$~$ \\
In & 1.00 & 0.2604(2) & 0 & 1/2 & 0.0093(4) \\
$~$&$~$&$~$&$~$&$~$&$~$ \\
\hline
\end{tabular}
\end{center}
\end{table}

\begin{table}[htb]
 \begin{center}
 \caption{Atomic coordinates and equivalent isotropic displacement parameters
($\AA^2$) for the on-line-grown YbPtIn system. U(eq) is defined as
one third of the trace of the orthogonal U$^{ij}$ tensor. Space
group \textit{P $\overline{6}$ 2 m}, a $=~7.5486(12)~\AA$, c
$=~3.7617(7)~\AA$, R = 0.0217, R$_\omega$ = 0.0453.} \label{t2}
\begin{tabular}{|c|c|c|c|c|c|} \hline
$~$&$~$&$~$&$~$&$~$&$~$ \\
& Occupancy & x & y & z & U(eq)\\
$~$&$~$&$~$&$~$&$~$&$~$ \\
\hline $~~~$&$~~~$&$~~~$&$~~~$&$~~~$&$~~~$ \\
$~~~$Yb$~~~$ & $~~~$1.00$~~~$& 0.4052(1) & $~~~$0$~~~$ & $~~~$0$~~~$ & 0.0076(2) \\
$~$&$~$&$~$&$~$&$~$&$~$ \\
Pt(1) & 0.986(9) & 0 & 0 & 0 & 0.0082(4) \\
$~$&$~$&$~$&$~$&$~$&$~$ \\
Pt(2) & 1.00 & 1/3 & 2/3 & 1/2 & 0.0071(2) \\
$~$&$~$&$~$&$~$&$~$&$~$ \\
In & 1.00 & 0.7378(2) & 0 & 1/2 & 0.0071(2) \\
$~$&$~$&$~$&$~$&$~$&$~$ \\
\hline
\end{tabular}
\end{center}
\end{table}

A high temperature factor was inferred for Pt(1) of the flux-grown
crystal during final stages of the refinement, which is usually
indicative of possible atomic deficiencies, symmetry reduction or
superstructure formation. No superstructure reflections were
observed, and symmetry reduction did not resolve the issue. However,
relaxing Pt(1) site occupancy resulted in a statistically
significant deficiency (0.06(1), Table \ref{t1}), and led to
improvements in the Pt(1) temperature factor and overall residual R
value. We also tested for possible deficiencies on the other atomic
sites in this crystal, but refined occupancies for Yb (1.00(6)),
Pt(2) (1.00(6)) and In (1.02(2)) did not suggest presence of atomic
deficiencies. Thus the composition of the flux-grown crystal can be
written as YbPt$_{0.98}$In, noting that the Pt$_{0.98}$ value
reflects the Pt stoichiometry of the whole unit cell, not just of
the Pt(1) site. In contrast to the flux-grown sample, the Pt(1)
temperature factor of the on-line grown crystal had a reasonable
value. Relaxing the Pt(1) occupancy yielded only a slight and
statistically insignificant deficiency of 0.014(9) (Table \ref{t2}).
Occupancy refinement for other atomic sites showed no deviations
from unity. Although small deficiencies on the Pt(1) site could not
be excluded, the YbPtIn formula is a good presentation of the
composition of the on-line crystal in terms of sensitivity of our
X-ray diffraction experiments. This difference in the stoichiometry
of the solution and on-line grown samples is consistent with YbPtIn
having a small width of formation extending towards the Pt-deficient
side; given that the initial melt composition is very Pt-poor
(\textit{i.e.}, Yb$_{0.4}$Pt$_{0.1}$In$_{0.5}$), it is expected to
be sensitive to such a small width of formation.

A closer look at the atoms' positions given in Tables
\ref{t1}-\ref{t2} suggests that two compounds might be mirror images
of each other; whereas racemic twinning could not be excluded for
either the solution or the on-line grown compounds, no evidence of
the existence of both "left" and "right" structures in each system
could be found.

The lattice parameters and unit cell volume for the solution-grown,
YbPt$_{0.98}$In crystals were $a~=~(7.55~\pm~0.01)$ \AA,
$c~=~(3.76~\pm~0.01)$ \AA $~~~$and $Vol~=~(186.28~\pm~0.51)$
\AA$^3$. The analogous unit cell dimensions on the on-line grown
crystals were slightly smaller: $a~=~(7.54~\pm~0.01)$ \AA,
$c~=~(3.75~\pm~0.01)$ \AA $~~~$and $Vol~=~(185.61~\pm~0.11)$
\AA$^3$.

\subsection{Low magnetic field comparison}

\begin{figure}
\begin{center}
\includegraphics[angle=0,width=90mm]{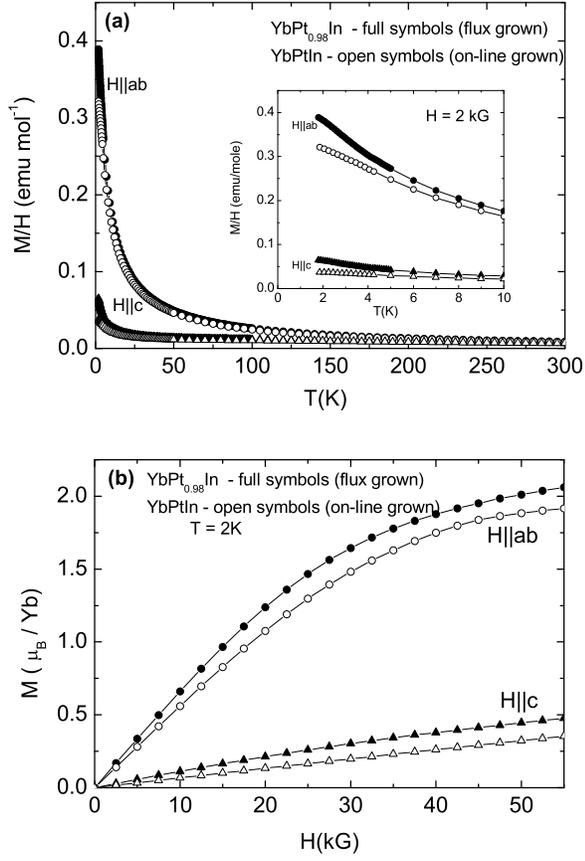}
\end{center}
\caption{(a) Anisotropic susceptibilities for H = 2 kG (with the
low-temperature part enlarged in the inset) and (b) field-dependent
magnetization at T = 2 K for YbPtIn (open symbols) and
YbPt$_{0.98}$In (full symbols).}\label{F01}
\end{figure}

Anisotropic magnetization measurements are presented in
Fig.\ref{F01} for both the YbPt$_{0.98}$In (full symbols) and the
YbPtIn (open symbols) compounds. As can be observed in
Fig.\ref{F01}a, the paramagnetic susceptibility indicates moderate
anisotropy for both systems (with $\chi_{ab}~/\chi_c~\sim~6$ at the
lowest temperature), and no clear sign of magnetic ordering down to
$T~=~1.8$ K. The anisotropic M(H) isotherms show a continuous
increase of the magnetization, with a trend towards saturation above
40 kG (Fig.\ref{F01}b) for H applied within the ab-plane; the axial
magnetization remains linear and significantly smaller than M$_{ab}$
up to our maximum field available for these measurements (55 kG).
Whereas qualitatively there is an overall similarity between the
corresponding data of the two compounds, the absolute values of both
susceptibility and field-dependent magnetization are slightly larger
for YbPt$_{0.98}$In than for YbPtIn. We believe the $\sim$ 10 to
20$\%$ difference to be too large to have been caused by weighing
errors alone, and thus we conclude that it may reflect the different
Kondo temperatures and exchange coupling due to the change of
stoichiometry in the two compounds.

\begin{figure}
\begin{center}
\includegraphics[angle=0,width=90mm]{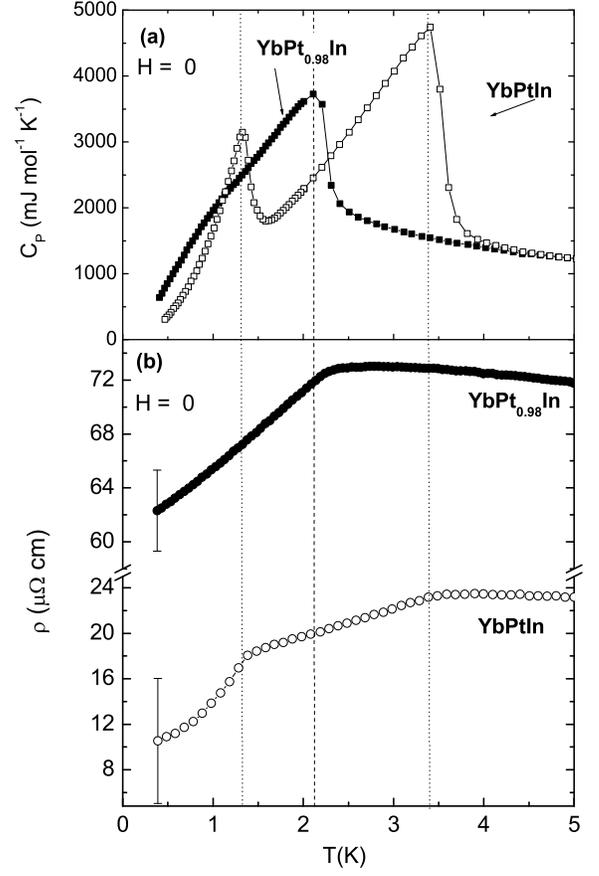}
\end{center}
\caption{(a) Low-temperature specific heat  and (b) resistivity data
for H = 0, for YbPtIn (open symbols) and YbPt$_{0.98}$In (full
symbols), with the resistivity error bars shown for the lowest
temperature (T = 0.4 K); the transition temperatures are indicated
by two dotted lines (for YbPtIn) and one dashed line (for
YbPt$_{0.98}$In). Note that the error bars in (b) are geometrical,
and thus they scale proportionally at higher
temperatures.}\label{F02}
\end{figure}

\begin{figure}
\begin{center}
\includegraphics[angle=0,width=90mm]{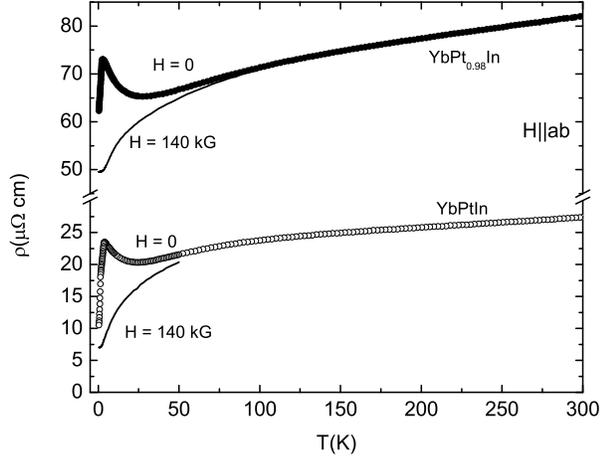}
\end{center}
\caption{Temperature-dependent resistivity for YbPt$_{0.98}$In and
YbPtIn, for H = 0 (symbols) and H = 140 kG (solid
lines).}\label{FRT300}
\end{figure}

The zero-field specific heat and resistivity data shown in
Fig.\ref{F02} are consistent with magnetic ordering in both
compounds, however at different temperatures: two well defined peaks
at T$~\approx~3.4$ K and 1.4 K are visible in the YbPtIn C$_P$(T)
data (open symbols, Fig.\ref{F02}a), whereas only one peak can be
distinguished, around 2 K, in the YbPt$_{0.98}$In data (full
symbols). These transition temperatures are marked by the vertical
dotted lines for the former compound, and by one dashed line for the
latter. It can be seen that the corresponding resistivity
measurements (Fig.\ref{F02}b) show changes in slope around the
respective transition temperatures. Another noticeable difference
between YbPtIn and YbPt$_{0.98}$In manifests in the resistivity
values (Fig.\ref{F02}b and \ref{FRT300}) with the ones for the
former compound being approximately three times smaller in the
latter one. The error bars shown for the lowest temperature $\rho$
values give a caliper of the uncertainty in estimating the
resistivity values for the two compounds, further confirming the
aforementioned difference. The larger residual resistivity in
YbPt$_{0.98}$In is consistent with the additional disorder
(\textit{i.e.}, site disorder) or presence of additional vacancies
in this type of crystals.

Thus the zero-field measurements indicate a dramatic effect of the
small Pt-deficiency on the ordered state in YbPtIn: the upper
transition is shifted down in temperature in the Pt-deficient
compound, whereas a second one is clearly identifiable only in
YbPtIn. In order to explore the differences between these two
samples more thoroughly, a systematic study of the field-dependence
of $\rho(T)$ and C$_p(T)$ was undertaken.

\subsection{High magnetic field measurements:
Y\lowercase{b}P\lowercase{t}I\lowercase{n}}

\textbf{$H~\parallel~ab$}\\

\begin{figure}
\includegraphics[angle=0,width=80mm]{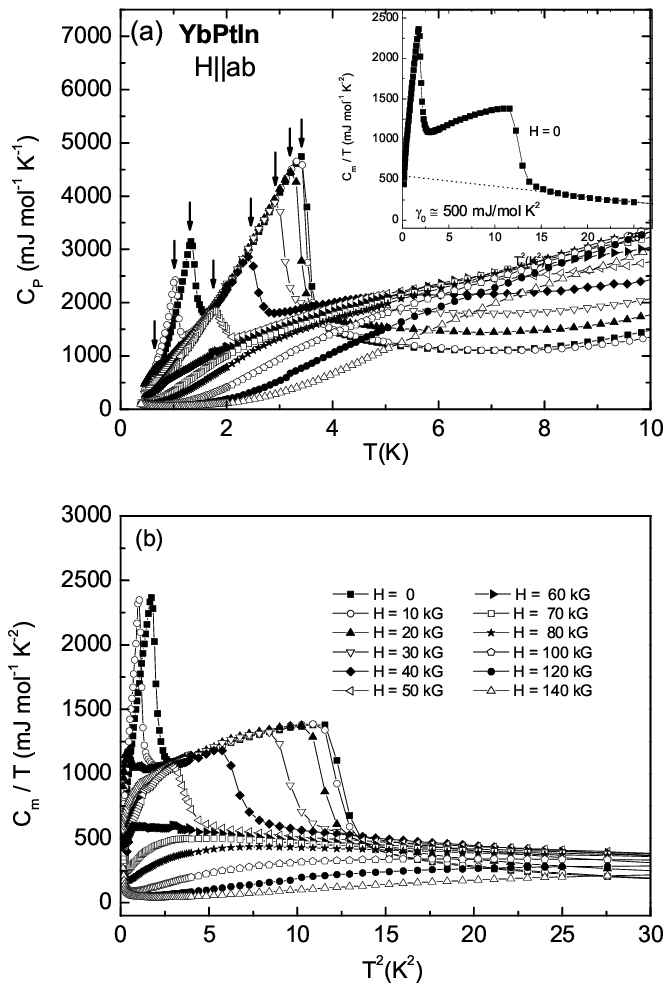}
\caption{(a) Low-temperature specific heat curves for YbPtIn, for H
$\parallel~ab$, with the arrows indicating the peak positions
associated with magnetic phase transitions; inset: the H = 0
C$_P/T$(T$^2$) curve, with its linear fit below 5 K (dotted line)
extrapolated down to T = 0 to provide a rough estimate of the
electronic specific heat coefficient $\gamma~\simeq~500$ mJ / mol
K$^2$ (b) low-temperature part of the C$_m/T$(T$^2$) curves for
various values of the applied field.}\label{F11}
\end{figure}

The low-temperature specific heat data for the on-line grown
compound YbPtIn is shown in Fig.\ref{F11}a, for various values of
the applied field. As already seen, two sharp peaks present in the H
= 0 data can be associated with magnetic phase transitions at T
$=~3.4$ K and 1.4 K. As the applied field is increased, these
transitions (indicated by small arrows in Fig.\ref{F11}a) move to
lower temperatures, and eventually drop below 0.4 K around 20 kG,
and 60 kG respectively. Trovarelli \textit{et al.}\cite{tro09} have
reported similar measurements up to 80 kG, which are consistent with
our data; however, their study did not include a systematic analysis
of the H - T phase diagram or the potential quantum critical
behavior in this compound. From the linear extrapolation of the
zero-field C$_P~/~T$ \textit{vs.} T$^2$ data from T $\sim$ 5 K down
to T = 0 (dotted line in the inset, Fig.\ref{F11}a), the electronic
specific heat coefficient $\gamma$ can be roughly estimated as
$\gamma~\simeq~500$ mJ / mol K$^2$. The magnetic component of the
specific heat is defined as $C_m~=~C_P$(YbPtIn)$~-~C_P$(LuPtIn), and
is shown in Fig.\ref{F11}b as $C_m~/~T$ \textit{vs.} T$^2$, for the
same field values as before.

\begin{figure}
\begin{center}
\includegraphics[angle=0,width=90mm]{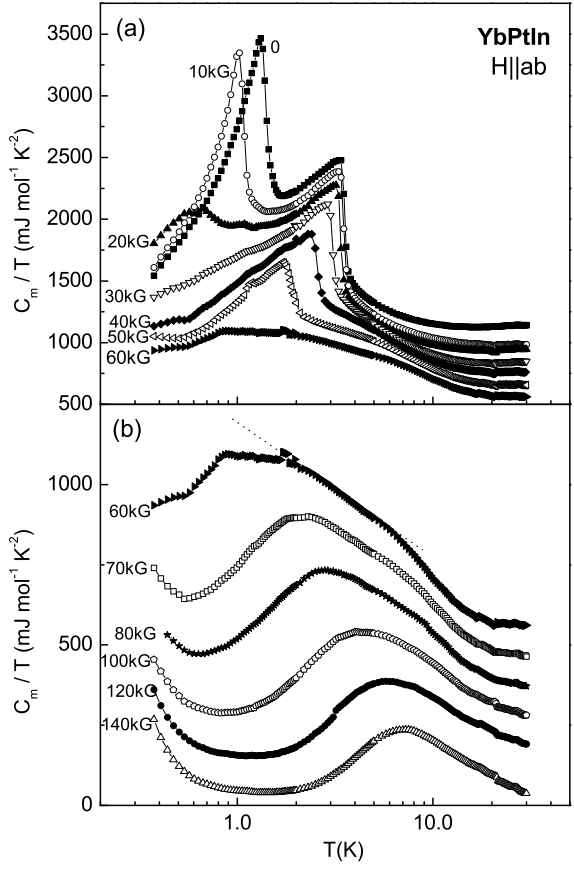}
\end{center}
\caption{Semi-log plot of C$_m/T$ \textit{vs.} T for (a) H = 0 - 60
kG and (b) H = 60 - 140 kG. All curves (except for the H = 140 kG
one) are shifted up by multiples of 100 mJ$/$mol K$^2$. The dotted
line (for H = 60 kG) is a guide to the eye for the largest region of
logarithmic divergency of C$_m/T$.}\label{F11a}
\end{figure}

When the magnetic specific heat is plotted in C$_m~/T$ vs. $\ln$ T
coordinates (Fig.\ref{F11a}), a reduced region of logarithmic
divergency (non-Fermi liquid NFL behavior) is apparent; however this
linear region in C$_m~/T(lnT)$ is more ambiguous than in other heavy
fermion compounds displaying NFL behavior (\textit{e.g.},
YbRh$_2$Si$_2$\cite{tro05} and YbAgGe\cite{bud11}). Because of a
downturn in the high field data (H $\geq$ 50 kG [Fig.\ref{F11a}b])
around 5 K for H = 50 kG, the largest logarithmic divergency which
occurs for H $\sim$ 60 kG, is limited to only a fraction of a decade
in temperature (1.5 K $<$ T $<$ 6.5 K). The above observations
suggests that a QCP may exist around a critical field value
H$^{ab}_c$ just above 60 kG, but the presence of a NFL region at
intermediate field values is less clearly defined than in the
previously studied Yb-based heavy fermion compounds.

One of the expressions considered in the scaling analysis at a
QCP\cite{tsv18} is the cross-over function $[C(H)~-~C(H=0)]~/~T$
\textit{vs.} $H/T^\beta$. In the case of YbRh$_2$Si$_2$\cite{pas09}
and YbAgGe\cite{bud12}, the $H/T^\beta$ range over which universal
scaling was observed in high fields corresponded to 1$/$T $<$ 3
K$^{-1}$, and 1.2 K$^{-1}$ respectively (with $\beta~>~1$). Due to
the slightly enhanced magnetic ordering temperature $T_{ord}~=~3.4$
K in YbPtIn, the analogous $1/$T range is drastically reduced (1$/$T
$<$ 0.3 K$^{-1}$), making the unambiguous determination of the
critical exponent $\beta$ essentially impossible.

\begin{figure}
\begin{center}
\includegraphics[angle=0,width=80mm]{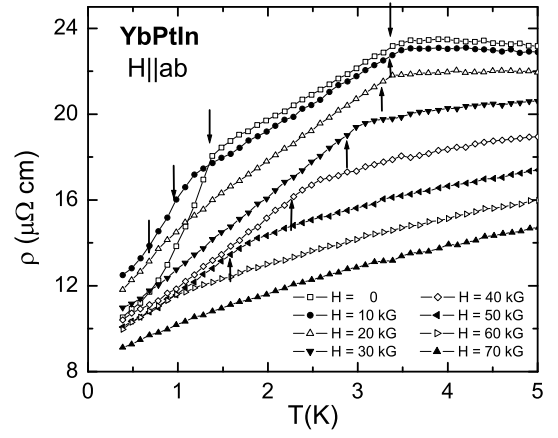}
\end{center}
\caption{Low-temperature resistivity data for YbPtIn taken at
different applied fields, for H $\parallel~ab$; the small arrows
indicate the magnetic transition temperatures}.\label{F12}
\end{figure}

Low-temperature resistivity curves for different values of the
applied magnetic field are shown in Fig.\ref{F12}. The
$\rho(T)\mid_H$ data are consistent with the presence of two
magnetic phase transitions at low fields; the small arrows indicate
these transition temperature values, as determined from maxima in
the d$\rho/dT$. Both these transitions are suppressed by increasing
applied field. Whereas the upper transition can still be detected
for H $<60$ kG, a field of about 20 kG is sufficient to drive the
lower one below our base temperature of 0.4 K. It is worth noting
that the critical field H$^{ab}_c~\sim~60$ kG, determined from the
$\rho$(T,H) data as the field required to suppress the magnetic
ordering, is close to the position of the QCP as inferred based on
the C$_P$ data (Fig.\ref{F11}-\ref{F11a}).

\begin{figure}
\begin{center}
\includegraphics[angle=0,width=80mm]{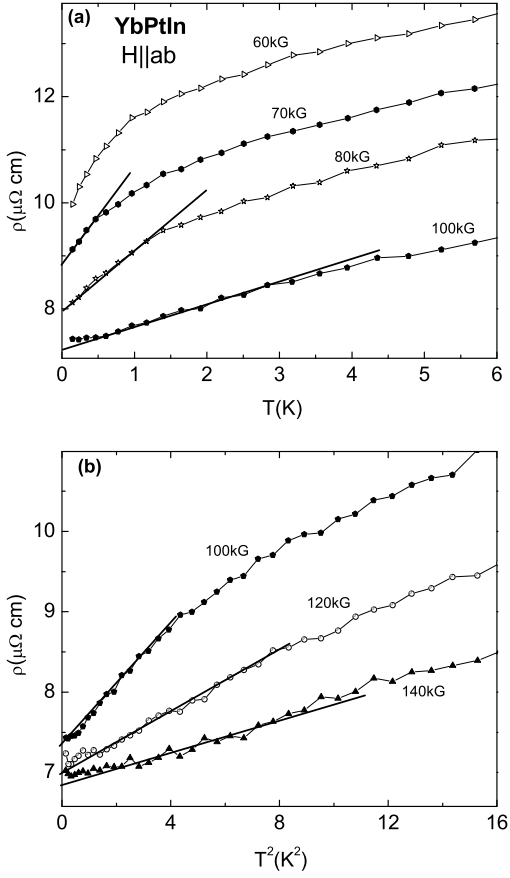}
\end{center} \caption{Low-temperature part of the (a) $\rho$(T) data
for H = 60, 70, 80 and 100 kG and (b) $\rho$(T$^2$) for H = 100, 120
and 140 kG.}\label{F12sqab}
\end{figure}

As Fig.\ref{F12sqab} indicates, when the magnetic field is increased
beyond H$^{ab}_c$, the temperature dependence of the resistivity is
ambiguous; particularly at fields just above H$^{ab}_c$ (H $\approx$
100 kG), it is difficult to distinguish between linear or quadratic
behavior of the resistivity as a function of temperature.
Consequently we leave the complete analysis of the
temperature-dependent resistivity outside the ordered state for the
discussion section.

\begin{figure}
\begin{center}
\includegraphics[angle=0,width=90mm]{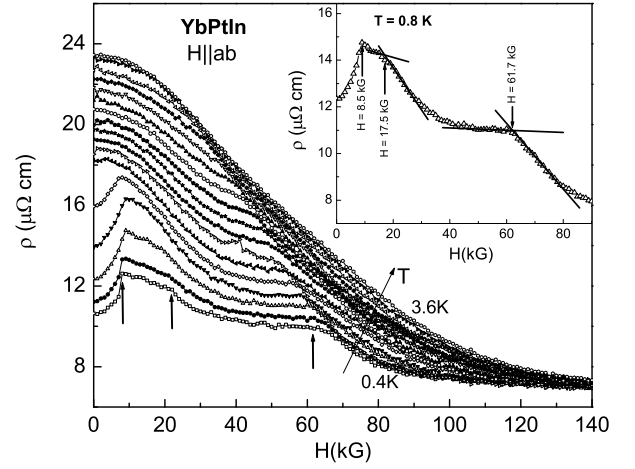}
\end{center}
\caption{$\rho$(H) isotherms for YbPtIn, for H $\parallel~ab$ and T
= 0.4 - 3.6 K, $\Delta$T = 0.2 K; the arrows point to the transition
fields at the lowest temperature, with the inset exemplifying how
these critical fields are determined for T = 0.8 K.}\label{F13}
\end{figure}

When magnetoresistance measurements ($\rho(H)\mid_T$) are performed
(Fig.\ref{F13}) three features are apparent at very low
temperatures, as indicated for T = 0.4 K by the small arrows. The
inset shows $\rho$(H) at T = 0.8 K, to exemplify how these
transition temperatures were inferred from these data. For T $\geq$
1 K, the two lower transitions merge and the resulting one is still
distinguishable up to approximately 1.4 K. The upper transition
moves down towards zero field as the temperature approaches
T$_N~=~3.4$ K.

\begin{figure}
\begin{center}
\includegraphics[angle=0,width=90mm]{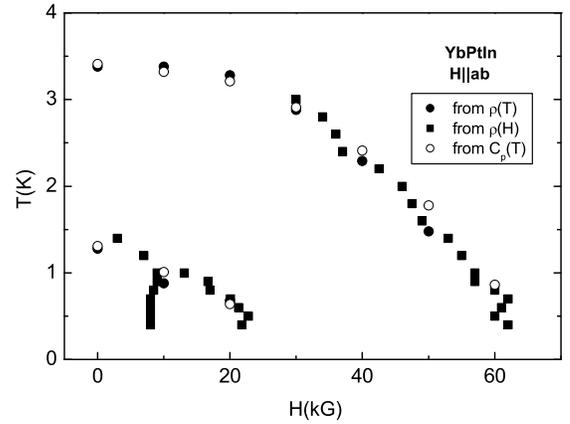}
\end{center}
\caption{H $\parallel~ab$ T - H phase diagram for YbPtIn, as
determined from resistivity (full symbols) and specific heat (open
symbols) data.}\label{F14}
\end{figure}

Based on the above field- and temperature-dependent thermodynamic
and transport measurements, a $T~-~H$ phase diagram for
$H\parallel~ab$ can be constructed (Fig.\ref{F14}): in zero magnetic
field, two magnetic phase transitions can be observed, around
T$_N~=~3.4$ K and T$_m~=~1.4$ K. Increasing magnetic field splits
the lower transition into two separate ones around H $\sim~10$ kG;
one of these phase lines drops towards our lowest temperature at
almost constant field, whereas the second one has a slower decrease
with field, such that it approaches T = 0.4 K around H = 20 kG. In a
similar manner, the upper transition is
driven down towards 0 around $H_c~=~60$ kG.\\

\textbf{$H~\parallel~c$}\\

\begin{figure}
\begin{center}
\includegraphics[angle=0,width=90mm]{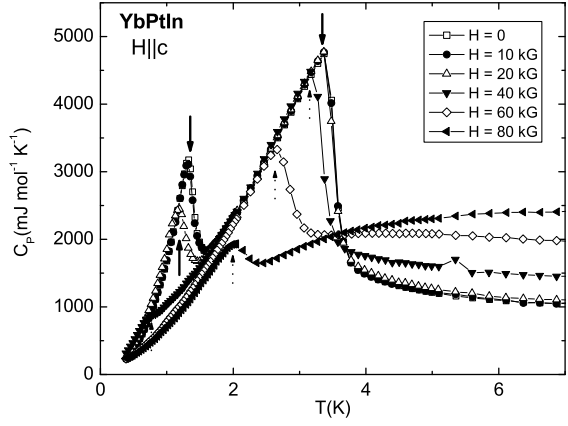}
\end{center}
\caption{$H~\parallel~c$ specific heat data for YbPtIn, for H = 0,
10, 20, 40, 60 and 80 kG; small arrows indicate the positions of
peaks possible associated with magnetic phase transitions (dotted
arrows: peaks on C$_P(T)$ data for possibly torqued sample [see
text]).}\label{F20}
\end{figure}

Fig.\ref{F20} presents specific heat data for YbPtIn, $H\parallel
c$, for fields up to 80 kG. High torques on this sample for this
orientation of the field prevented us from completing these
measurements up to 140 kG. Moreover, as will be shown below, there
are significant discrepancies between the transition temperatures
determined even from the intermediate-field specific heat data and
transport measurements (\textit{i.e.}, for H $\geq$ 40 kG). This
observation prompts us to suspect that significant torques may have
changed the sample orientation for the specific heat measurements,
even for fields significantly lower than 80 kG.

In zero field, we can confirm the two magnetic transitions observed
before, at T$_N~=~3.4$ K, and T$_m~=~1.4$ K respectively; as the
small arrows indicate, the lower-T transition is driven down in
field, and falls below 0.4 K for H $~>~40$ kG, whereas the upper
transition persists above 80 kG.

\begin{figure}
\begin{center}
\includegraphics[angle=0,width=90mm]{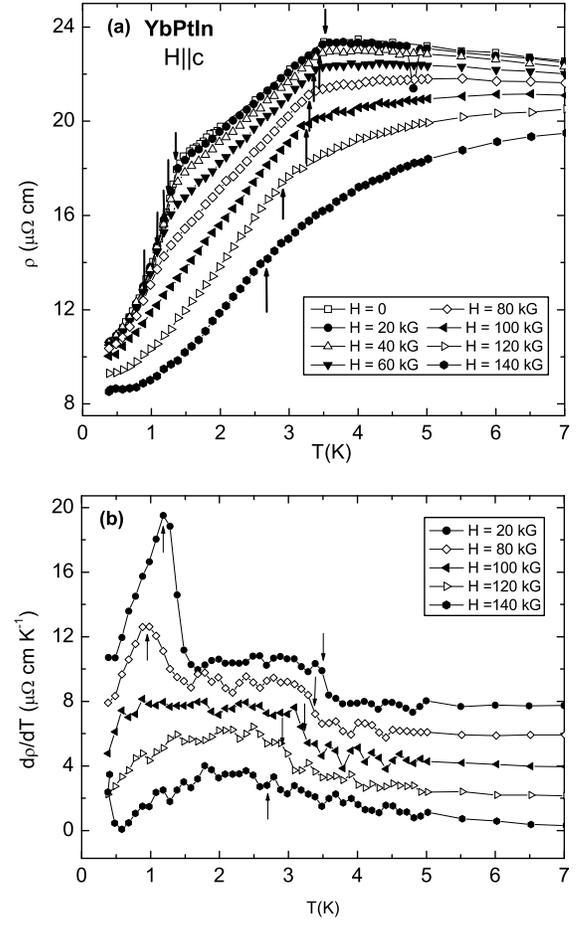}
\end{center}
\caption{(a) Low-temperature $\rho$(T) data for YbPtIn taken at
different applied fields, for H $\parallel~c$. (b) $d\rho/dT$ curves
for H = 20, 80, 100, 120 and 140 kG. On both plots, the small arrows
indicate the magnetic transition temperatures.}\label{F21}
\end{figure}

\begin{figure}
\begin{center}
\includegraphics[angle=0,width=90mm]{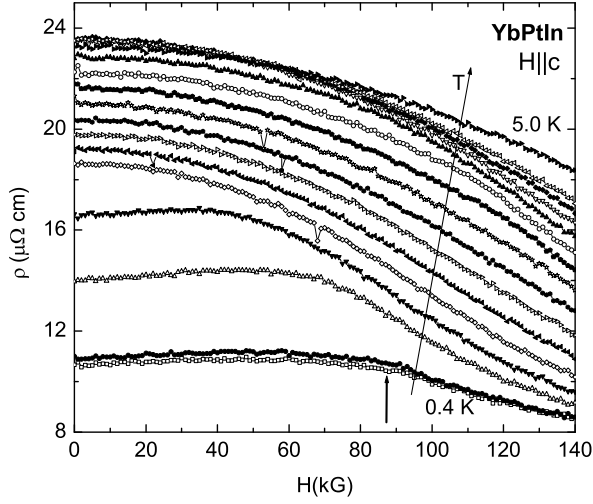}
\end{center}
\caption{$\rho$(H) isotherms for T = 0.4, 0.5, 1 - 4 K ($\Delta$T =
0.25 K) and 5 K. Small arrow indicates the critical field position
at the lowest temperature (T = 0.4 K).}\label{F21rh}
\end{figure}

The temperature and field dependent resistivity data
(Fig.\ref{F21}-\ref{F21rh}) indicate a much slower suppression of
the magnetic order with the applied field. In Fig.\ref{F21}a,
$\rho(T)$ curves are shown, with the arrows indicating the
transition temperatures as determined from $d\rho/dT$.
Fig.\ref{F21}b presents a subset of these derivatives, to illustrate
the criteria for determining the temperatures: for the lower
transition, a peak in $d\rho/dT$ broadens as field is increased, and
disappears for H $>$ 80 kG; the upper transition is marked by a step
in these derivatives, which also broadens as H increases. At the
highest measured field (\textit{i.e.}, 140 kG) we are unable to
distinguish between a very broad step (with a possible transition
temperature marked by the small arrow) or a cross-over in
corresponding $d\rho/dT$. The field-dependent resistivity data
(Fig.\ref{F21rh}) are indicative of a low temperature transition
consistent with that seen in $\rho(T)$, with the critical fields
determined from onsets.

Given the above C$_P$(T,H) and $\rho$(T,H) data, we suspect that
magnetic fields H $>$ 20 kG deform the four wires supporting the
He-3 specific heat platform used for the C$_P$(T,H) measurements,
whereas the resistivity sample appears to be well held in place by
grease on the rigid platform. Consequently, at high fields, the two
sets of data (C$_P$(T,H) and $\rho$(T,H)) may not correspond to the
same orientation of the field ($H~\parallel~c$), yielding different
transition temperature values for the corresponding applied fields.

\begin{figure}
\begin{center}
\includegraphics[angle=0,width=90mm]{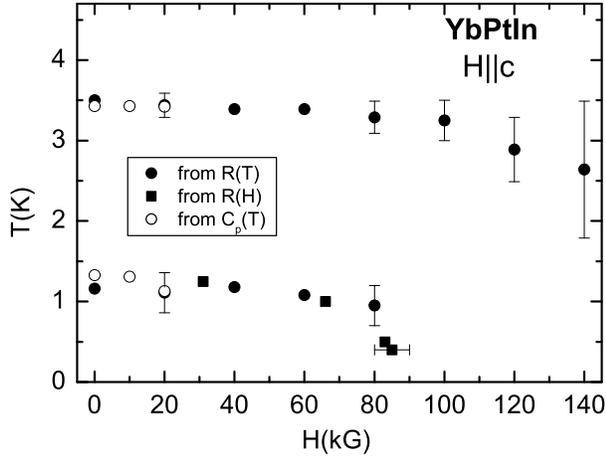}
\end{center}
\caption{H $\parallel c$ T - H phase diagram for YbPtIn, as
determined from resistivity (full symbols) and specific heat (open
symbols) data. Error bars on points from $\rho$(T) data at H = 20,
80, 100, 120 and 140 kG, and from $\rho$(H) at T = 0.4 K shown as a
caliper of the errors in determining the points on this phase
diagram.}\label{F22}
\end{figure}

As a result, in constructing the T - H phase diagram for H$\parallel
c$ (Fig.\ref{F22}), we will consider the T$_c$ values as determined
from the $\rho$(T,H) data up to H = 140 kG, and only the H $\leq$ 20
kG ones based on specific heat measurements. Also shown are error
bars for points determined from $\rho$(T) data at several field
values (\textit{i.e.}, for H = 20, 80, 100, 120 and 140 kG), and for
the point obtained from $\rho$(H) at our minimum temperature (T =
0.4 K); these give a caliper of the errors bars in determining the
points on this phase diagram for the whole field and temperature
range. Two transitions can be observed in Fig.\ref{F22}, at low
fields, around 3.4 K, and 1.4 K respectively. As H is being
increased, the low temperature line slowly approaches T = 0 around H
$~\sim~85$ kG. The step in $d\rho/dT$ associated with the upper
transition (Fig.\ref{F21}b) broadens as the field increases,
resulting in increasingly large error bars in determining these
transition temperatures. As already mentioned, it is uncertain if
the transition persists up to H = 140 kG, or if cross-over occurs
between 120 and 140 kG.

\subsection{High magnetic field measurements:
Y\lowercase{b}P\lowercase{t}$_{0.98}$I\lowercase{n}}

\textbf{$H~\parallel~ab$}\\

Given the differences between YbPtIn and YbPt$_{0.98}$In evidenced
by both thermodynamic and transport data
(Figs.\ref{F01}-\ref{FRT300}), it is desirable to compare similar
measurements on the two compounds, and to study the effect of the
small stoichiometry change on the field-induced QCP.

\begin{figure}
\begin{center}
\includegraphics[angle=0,width=90mm]{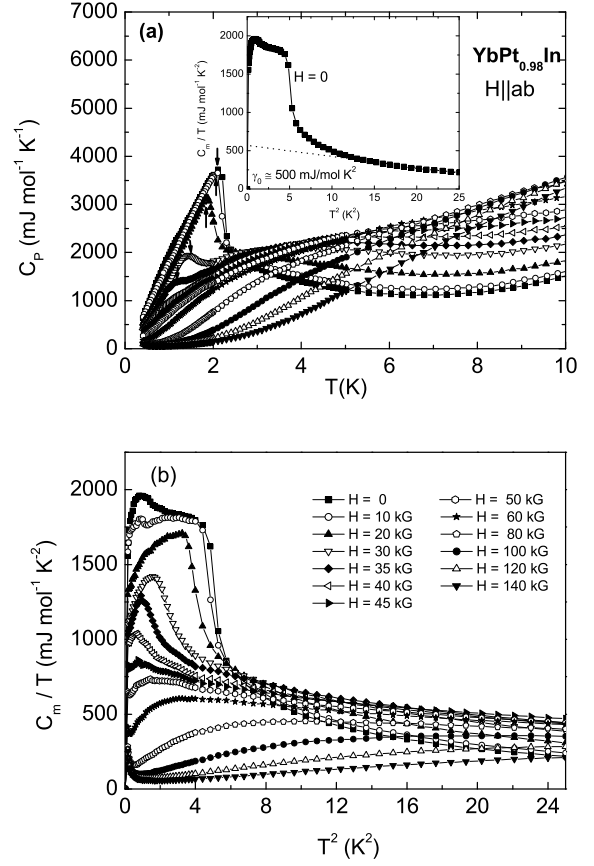}
\end{center}
\caption{(a) Low-temperature specific heat curves for
YbPt$_{0.98}$In, for H $\parallel~ab$, with the arrows indicating
the positions of the peaks associated with magnetic ordering; inset:
the H = 0 C$_P/T$(T$^2$) curve, with its linear fit below 5 K
(dotted line) extrapolated down to T = 0 to provide a rough estimate
of the electronic specific heat coefficient $\gamma~\simeq~500$ mJ /
mol K$^2$ (b) low-temperature part of the C$_m/T$(T$^2$) curves for
various values of the applied field.}\label{F03}
\end{figure}

Consequently, in Fig.\ref{F03} we present the low-temperature
specific heat data of YbPt$_{0.98}$In, for various values of the
applied field H $\parallel~ab$. A well-defined peak at T $=~2.1$ K
in the H $=~0$ data may be associated with the magnetic ordering of
this compound. As the applied magnetic field is increased, this
transition (indicated by small arrows) moves to lower temperatures,
and eventually drops below 0.4 K around 35 kG. From the linear
extrapolation of the H = 0 data from T $\sim~5$ K down to T = 0
(inset, Fig.\ref{F03}a), we can estimate the electronic specific
heat coefficient $\gamma$ as $\gamma~\approx~500$ mJ / mol K$^2$.
Fig.\ref{F03}b the magnetic specific heat data for the same field
values, $C_m~=~C_P$(YbPt$_{0.98}$In)$~-~C_P$(LuPtIn), plotted as
C$_m/T$ \textit{vs.} T$^2$.

Fig.\ref{F03b} shows the magnetic specific heat as $C_m/T$
\textit{vs.} $\ln~T$, for the same field values as before. A
logarithmic divergence can be observed in these data, with the
largest temperature region where $C_m/T(lnT)$ occurring around H =
35 kG (dotted line). However, on the next measured curve (\textit{i.
e.}, for H = 40 kG) the linear region extends over a comparable
temperature interval, at slightly higher temperatures than in the H
= 35 kG case. It thus appears that the largest temperature region
(close to a decade) for the logarithmic divergency of the C$_m/T$
data may occur for some intermediate field value (35 kG $<$ H $<$ 40
kG). These data could be described as $C_m/T~=~\gamma'_0\ln(T_0/T)$,
with the ranges for $\gamma'_0$ and T$_0$ determined from the linear
fits on the H = 35 and 40 kG curves: 420 mJ$/$mol
K$^2~<~\gamma'_0~<$ 430 mJ$/$mol K$^2$, and 14.7 K $>$ T$_0~>$ 13.5
K. The above observations seem consistent with a QCP in this
compound with critical field just above 35 kG.

\begin{figure}
\begin{center}
\includegraphics[angle=0,width=88mm]{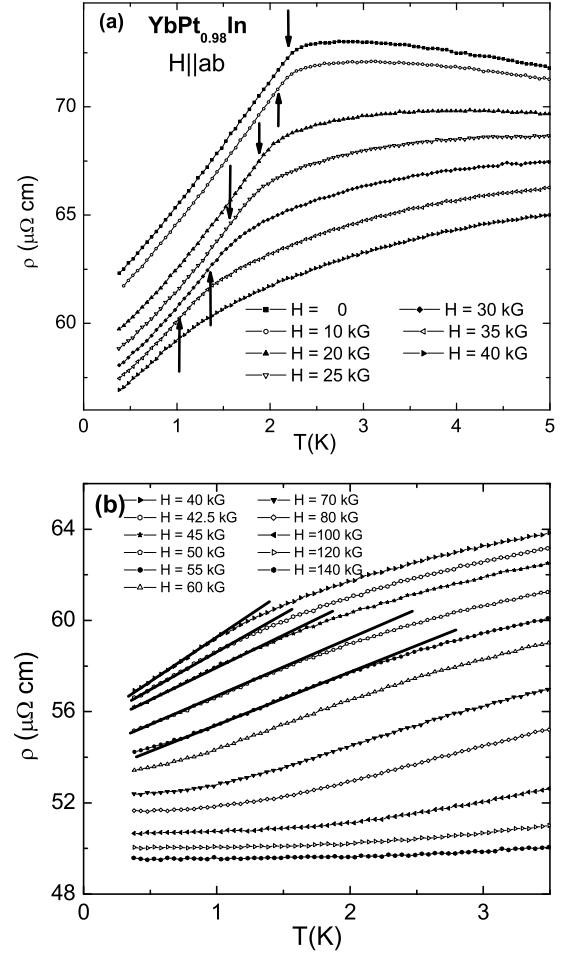}
\end{center}
\caption{YbPt$_{0.98}$In low-temperature resistivity data for H
$\parallel~ab$, for (a) H = 0, 10, 20, 25, 30, 35 and 40 kG and (b)
H = 40, 42.5, 45, 50, 55, 60, 70, 80, 100, 120 and 140 kG. Small
arrows in (a) indicate the possible ordering temperatures, whereas
in (b) the solid lines are guides to the eye for the potentially
linear regions on the H = 42.5, 45, 50 and 55 kG curves.}\label{F04}
\end{figure}

Fig.\ref{F04} shows the low-temperature resistivity data for various
values of the applied magnetic field. A maximum in d$\rho/dT$,
associated with the magnetic ordering can be identified at T
$\approx~2.2$ K in the H = 0 data, and is marked by small arrow in
Fig.\ref{F04}a; as the field is increased above 35 kG, small arrows
indicate that this transition temperature drops below 0.4 K ,
consistent with the specific heat data.

\begin{figure}
\begin{center}
\includegraphics[angle=0,width=82mm]{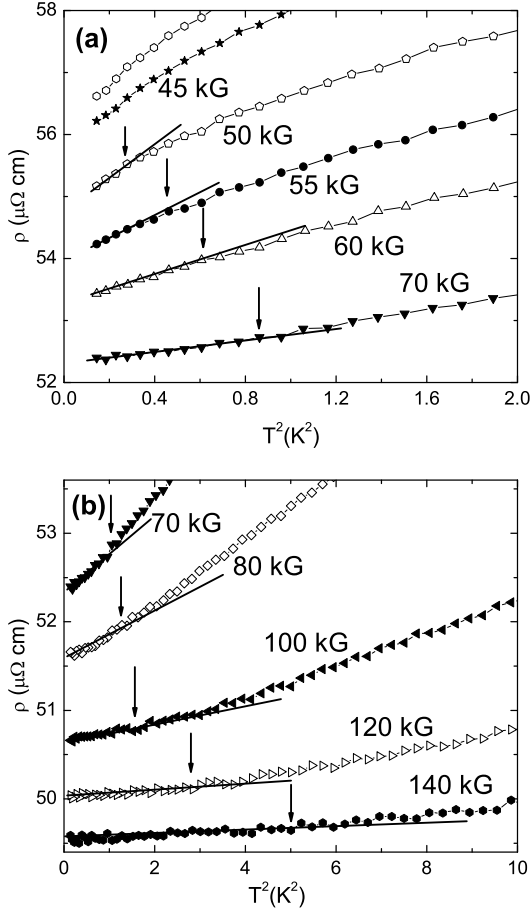}
\end{center}
\caption{Low-temperature part of the YbPt$_{0.98}$In $\rho$(T$^2$)
data, for (a) H = 42.5, 45, 50, 55, 60 and 70 kG and (b) H = 70, 80,
100, 120 and 140 kG, with solid lines as guides to the eye for the
potentially linear regions for the H $\geq$ 45 kG
curves.}\label{F04sqab}
\end{figure}

As was the case of YbPtIn, in YbPt$_{0.98}$In the temperature
dependence of the resistivity outside the ordered state is
ambiguous. From figs.\ref{F04}b-\ref{F04sqab} it is unclear whether
the resistivity is linear or quadratic in field for H $>~H^{ab}_c$,
particularly for intermediate field values (H $\leq$ 55 kG). A
detailed analysis of the resistivity data above the critical field
will be performed in the discussion section, for both compounds.
This should allow us to better clarify the position of the QCP and
the existence of a Fermi liquid-like (FL) regime in these systems.

\begin{figure}
\begin{center}
\includegraphics[angle=0,width=90mm]{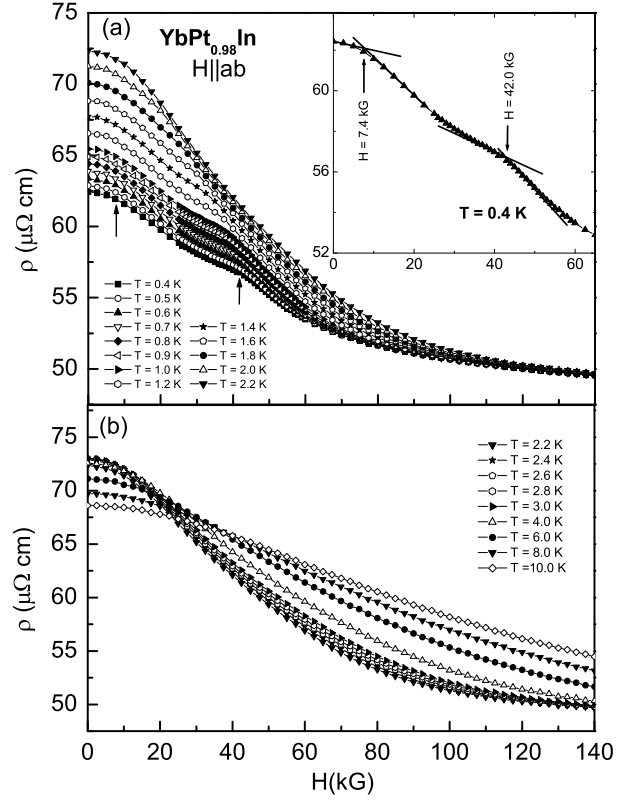}
\end{center}
\caption{$\rho$(H) isotherms for YbPt$_{0.98}$In, for H
$\parallel~ab$ and (a) T = 0.4 - 1.0 K ($\Delta$T = 0.1 K), 1.2 -
2.2 K ($\Delta$T = 0.2 K), and (b) T = 2.2 - 3.0 K ($\Delta$T = 0.2
K), 4, 6, 8 and 10 K; the arrows point to the transition fields at T
= 0.4 K, with the inset exemplifying how these critical fields were
determined.}\label{F05}
\end{figure}

Transverse magnetoresistance measurements were taken at constant
temperatures ranging from 0.4 K to 10 K. As shown in Fig.\ref{F05},
two different temperature regimes can be identified in these data:
for T $=~0.4~-~1.8$ K (Fig.\ref{F05}a), two transitions can be
distinguished. The small arrows mark the positions for these two
transitions for T = 0.4 K, whereas the inset illustrates how these
critical field values were determined. As the temperature increases
up to about 1.8 K, the upper transition moves down in field and
broadens, whereas the position of the lower one seems almost
unaffected by the change in temperature. For temperatures higher
than 2 K (Fig.\ref{F05}b), the magnetoresistance isotherms display
only a broad feature that looks more like a cross-over rather than a
transition.

\begin{figure}
\begin{center}
\includegraphics[angle=0,width=90mm]{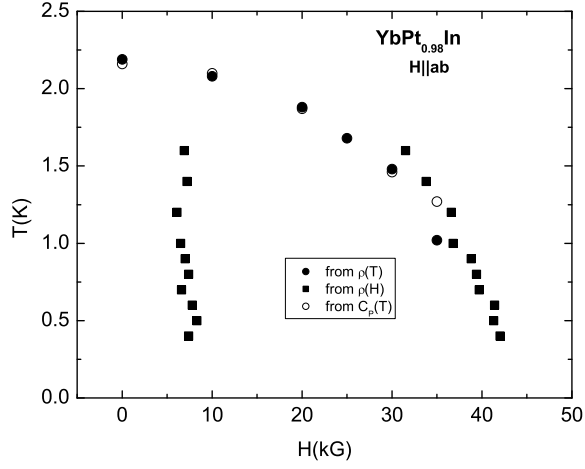}
\end{center}
\caption{H $\parallel~ab$ phase diagram for YbPt$_{0.98}$In, as
determined based on specific heat (open symbols) and transport (full
symbols) data. The error bars on the high-field points represent the
uncertainty in determining the cross-over temperatures.}\label{F06}
\end{figure}

Using the detailed C$_P$(T,H) and $\rho$(T,H) measurements discussed
above, the YbPt$_{0.98}$In T - H phase diagram for H$~\parallel~ab$
can be constructed. As can be seen in Fig.\ref{F06}, it is
qualitatively similar to the corresponding T - H phase diagram for
the stoichiometric compound (Fig.\ref{F14}): in YbPt$_{0.98}$In
magnetic ordering occurs at around 2.2 K. An almost field
independent phase line is apparent in the $\rho(H)\mid_T$ data
around 8 kG, and it appears to persist close to the magnetic
ordering temperature. Increasing applied field drives the higher
transition towards T = 0 at a critical field values around 35 kG.\\

\textbf{$H~\parallel~c$}\\

\begin{figure}
\begin{center}
\includegraphics[angle=0,width=90mm]{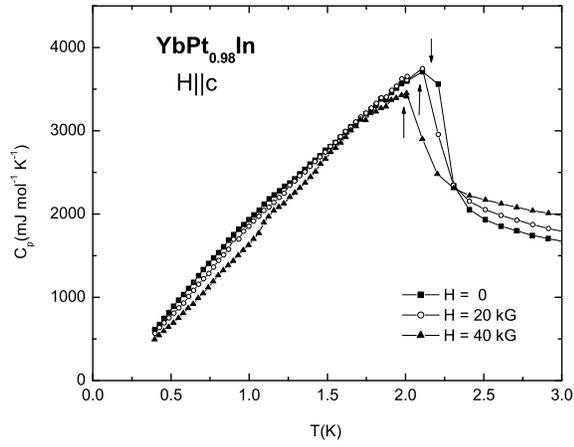}
\end{center}
\caption{$H~\parallel~c$ specific heat data for YbPt$_{0.98}$In, for
H = 0,20 and 40 kG; small arrows indicate the positions of peaks
possibly associated with the magnetic phase transition.}\label{F17}
\end{figure}

The similarities observed previously between YbPtIn and
YbPt$_{0.98}$In are also present for the $H~\parallel~c$ direction,
as the C$_P$(T,H) and $\rho$(T,H) measurements indicate. As in the
case of the stoichiometric compound, significant torques on the
specific heat platform and sample for H $\geq$ 50 kG may be the
cause of the different transition temperature values, as determined
by the two data sets mentioned above. Therefore we will only take
into consideration C$_P$(T,H) data for H $<$ 50 kG (Fig.\ref{F17}).
Similar to the $H~\parallel~ab$ measurements (Fig.\ref{F03}), the
$H~\parallel~c$ C$_P$ curves reveal a magnetic transition around 2.1
K for H = 0 (Fig.\ref{F17}), which drops to $\sim$ 2 K for H = 40
kG, before the sample torques significantly; small arrows indicate
the position of the transition temperature for the three curves
shown in Fig.\ref{F17}.

\begin{figure}
\begin{center}
\includegraphics[angle=0,width=90mm]{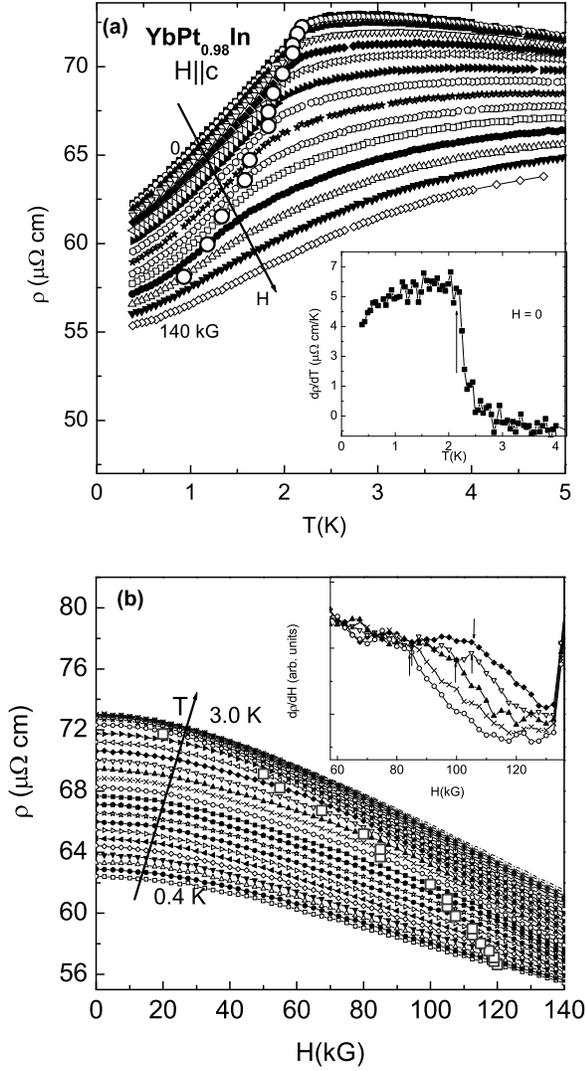}
\end{center}
\caption{(a) YbPt$_{0.98}$In low-temperature resistivity data for H
$\parallel~c$, for H = 0 - 140 kG ($\Delta$H = 10 kG), with the
large dots marking the possible magnetic ordering temperature;
inset: H = 0 d$\rho/$dT, with the small arrow marking the transition
temperature. (b) $\rho$(H) isotherms for T = 0.4 - 3.0 K ($\Delta$T
= 0.1 K). The critical fields for the possible phase transition were
determined as local maxima in d$\rho/dH$ (as illustrated in the
inset for T = 1.2 - 1.6 K), and are shown as large
squares.}\label{F18}
\end{figure}

The temperature and field dependent resistivity data (Fig.\ref{F18})
indicate a slow suppression of the magnetic order with the applied
field. Fig.\ref{F18}a shows the $\rho$(T) curves in various applied
fields, with the large circles marking the phase transition as
determined from d$\rho/dT$; the inset illustrates how the transition
temperature was determined for H = 0. The critical field required to
suppress this transition below our base temperature appears to be
around 120 kG. Given the limited temperature range at these high
fields, for H = 130 and 140 kG we were unable to distinguish a
linear or quadratic temperature dependence of the resistivity. The
magnetoresistance isotherms are presented in Fig.\ref{F18}b, and the
large squares on this plot also indicate the high-T magnetic phase
transition; in the inset, a few d$\rho/dH$ curves are shown to
illustrate how the critical field values for the transition were
determined.

\begin{figure}
\begin{center}
\includegraphics[angle=0,width=90mm]{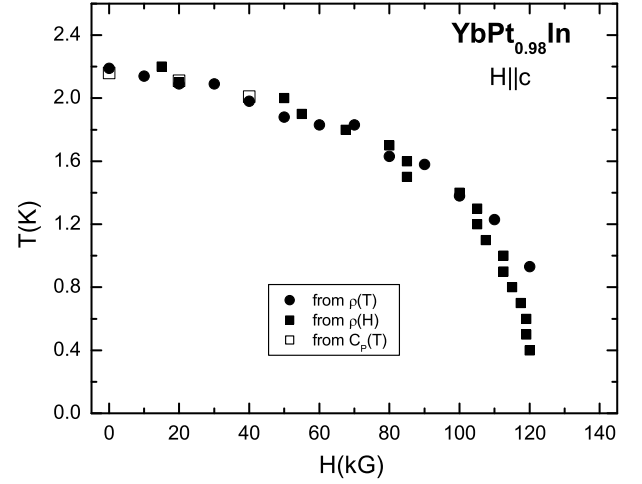}
\end{center}
\caption{T - H phase diagram for YbPt$_{0.98}$In ($H~\parallel~c$),
as determined from the specific heat (open symbols) and resistivity
(full symbols) measurements.}\label{F19}
\end{figure}

Based on the C$_P$(T,H) and $\rho$(T,H) presented above, the
$H~\parallel~c$ T - H phase diagram for YbPt$_{0.98}$In can be
obtained, as shown in Fig.\ref{F19}. At low fields, this T - H phase
diagram is consistent with the in-plane one for this compound: a
magnetic transition is apparent around 2.2 K, but the possible
second one around 1.0 K is not visible in the H$\parallel$c
measurements; the T $\simeq$ 2.2 K transition is driven down in
temperature by increasing applied fields, and it approaches our base
temperature (\textit{i.e.}, 0.4 K) around 120 kG. Lack of
measurements below 0.4 K or above 140 kG limits our ability to probe
the existence of a QCP in this orientation, similar to the one at
H$^{ab}_c~=~35$ kG for $H~\parallel~ab$.

\section{discussion}

For both the YbPtIn and YbPt$_{0.98}$In compounds a number of
similar properties, as well as systematic differences, could be
distinguished. First, the small stoichiometry difference was
apparent from single crystal x-ray measurements, which also resulted
in slightly reduced lattice parameters and unit cell volumes in the
YbPtIn system. Furthermore, the resistivity values were shifted
towards higher values in the Pt-deficient compound
(Figs.\ref{F02}b-\ref{FRT300}) for the whole temperature (T =
0.4-300 K) and field ranges (H = 0-140 kG) of our measurements. High
enough fields suppressed the magnetic order in both systems, at
least for field applied within the basal plane; in a similar manner
to the T-scale in these compounds, the critical field value was
reduced in YbPt$_{0.98}$In by comparison to the analogous one in
YbPtIn. For $H \parallel c$, the field values required to suppress
the magnetic order in YbPt$_{0.98}$In and YbPtIn were close to, or
respectively higher than our maximum available field (\textit{i.e.},
140 kG); this precluded us from studying the low temperature
properties of the two compounds outside the ordered state, for this
orientation of the field.

For $H~\parallel~ab$ NFL-like behavior seems to occur in
YbPt$_{0.98}$In for intermediate field values above the ordering
temperatures, as indicated by the logarithmic divergence of the
C$_m/T$ data (Fig.\ref{F03b}a). In YbPtIn the NFL region is less
clearly defined in the specific heat, as its logarithmic divergence
was limited to a small temperature range by a downturn in the
C$_m/T(\ln T)$ data (Fig.\ref{F11a}) towards high T. Even more
ambiguous was the temperature dependence of the resistivity data in
both compounds, for H $>~H^{ab}_c$. Below we are presenting a
detailed analysis of these data, aimed at distinguishing between, if
existent, the NFL and FL regimes, and providing a more accurate
estimate of the QCP position in each of the two compounds.

\subsection{H$\parallel$ab resistivity data above the magnetically ordered state}

Given the ambiguity already mentioned in the power law that best
describes $\rho(T)$ data above the ordered state, we attempt to
determine the exponent $\beta$ when the temperature dependence of
the resistivity is given by: $\Delta \rho(T)~=~A~T^\beta$, where
$\Delta \rho(T)~=~\rho(T)~-~\rho_0$. This is best done by
determining the slope of the $\Delta \rho(T)$ data on log-log plots
at various field values. However, the accuracy of the resulting
$\beta(H)$ values is highly dependent on accurate estimates of the
residual resistivity values $\rho_0(H)$. In turn, the estimates of
the $\rho_0(H)$ values are made difficult by their enhanced values
compared to the over-all resistivity values (fig.\ref{FRT300}).

\begin{figure}
\begin{center}
\includegraphics[angle=0,width=88mm]{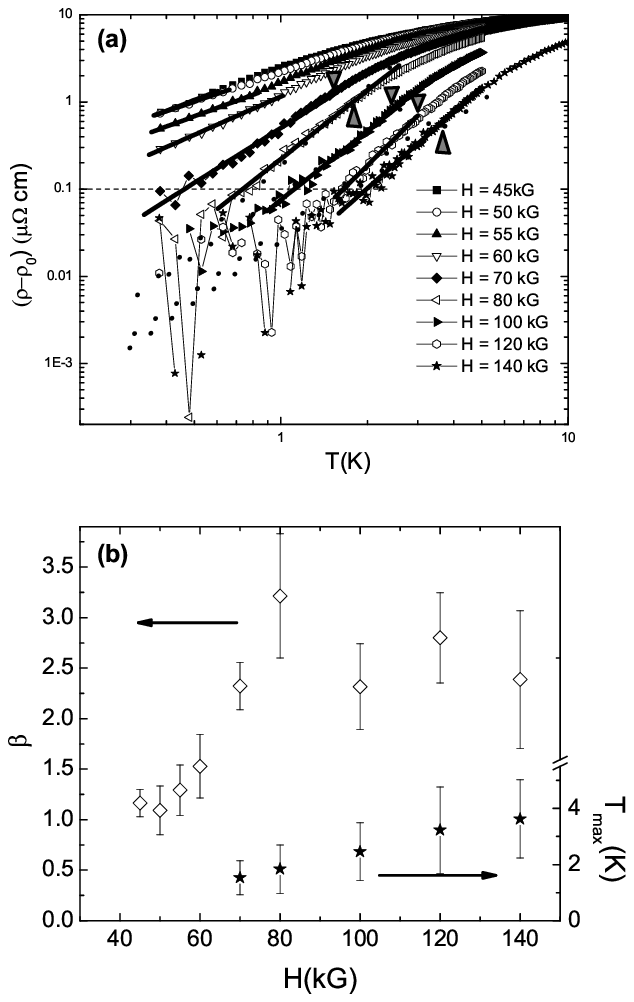}
\end{center}
\caption{(a) Low-temperature part of the YbPt$_{0.98}$In $\ln(\Delta
\rho)$ \textit{vs.} $\ln T$ data for H = 45, 50, 55, 60, 70, 80,
100, 120 and 140 kG, with their respective linear fits at very low T
(solid lines). The large triangles indicate the temperatures at
which deviations from linearity occur. (b) Exponent $\beta$ (left)
and cross-over temperature T$_{max}$ (right) as a function of H,
with the corresponding error bars.}\label{F04sqcd}
\end{figure}

\begin{figure}
\begin{center}
\includegraphics[angle=0,width=88mm]{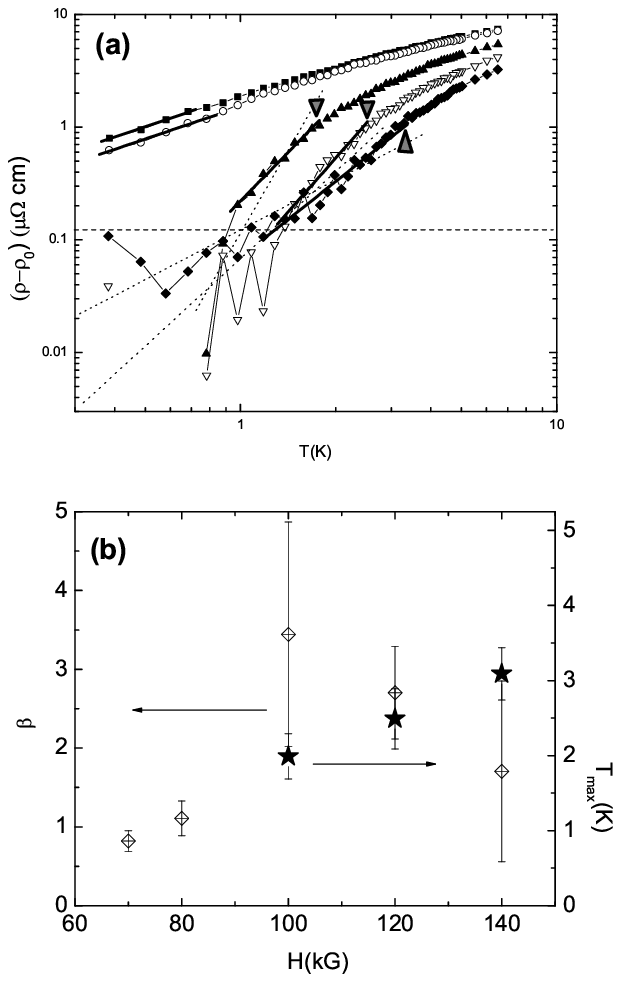}
\end{center}
\caption{(a) Low-temperature part of the YbPtIn $\ln(\Delta \rho)$
\textit{vs.} $\ln T$ data for H = 70, 80, 100, 120 and 140 kG, with
their respective linear fits at very low T (solid lines). The large
triangles indicate the temperatures at which deviations from
linearity occur. (b) Exponent $\beta$ (left) and cross-over
temperature T$_{max}$ (right) as a function of H, with their
corresponding error bars.}\label{F12sqcd}
\end{figure}

Consequently, we determine the residual resistivity values using two
criteria: firstly, from fits of the low-T resistivity data to a
$\rho_0~+~A~T^\beta$ function; secondly, given that
figs.\ref{F04}b-\ref{F04sqab} (for YbPt$_{0.98}$In), or
fig.\ref{F12sqab} (for YbPtIn) suggests that low-T $\rho(T)$ data
are close to being either linear or quadratic in T, we also choose
(for mathematical rather than physical reasons) a second-order
polynomial function to determine $\rho_0$ as a function of field:
$\rho(T)~=~\rho_0~+~A~T~+~B~T^2$. For each field value, the average
$\rho_0$ resulting from the two aforementioned criteria is used to
calculate $\Delta \rho(T)~=~\rho(T)~-~\rho_0$ plotted in
Fig.\ref{F04sqcd}a and \ref{F12sqcd}a respectively, on a log-log
scale. The large data scattering towards the lowest temperatures,
particularly at high fields, is due to the fact that $\Delta~\rho$
is calculated as the difference between two comparably large values,
$\rho(T)$ and $\rho_0$. As a consequence, the values of the exponent
$\beta$ are being determined from (i) linear fits shown as solid
lines, for $\Delta \rho(T)~=~\rho(T)~-~\rho_0$ larger than $\sim$
0.1 $\mu \Omega~cm$ (marked by the dashed horizontal line), and (ii)
from similar linear fits down to the lowest measured temperature
(dotted lines). Together with the exponent values originally
determined from fitting the low-T data to $\rho_0~+~A~T^\beta$
functions, we can now determine the $\beta$ values shown in
Fig.\ref{F04sqcd}b and \ref{F12sqcd}b, as full symbols, with their
corresponding error bars.

In the case of YbPt$_{0.98}$In (Fig.\ref{F04sqcd}a), the resistivity
is linear on the log-log scale, up to some maximum temperatures
T$_{max}$ marked with large triangles, with a large change in slope
for 60 kG $<$ H $<$ 70 kG. This is equivalent to the exponent
$\beta$ having a discontinuous change from $\beta~\approx~1$ below
60 kG (\textit{i.e.,} for H $>$ H$^{ab}_c$), to $\beta~\approx~2$ in
higher fields (H $\gg$ H$^{ab}_c$), as can be seen in
Fig.\ref{F04sqcd}b. Also shown in Fig.\ref{F04sqcd}b are the
temperatures T$_{max}$ below which the low-temperature $\Delta \rho$
manifests a Fermi liquid-like exponent of $\beta~\approx$ 2,
analogous to the cross-over previously reported for
YbRh$_2$Si$_2$\cite{tro05} and YbAgGe\cite{bud11}. The departure
from linearity on the resistivity plots can be determined up to a
temperature range, and not a unique temperature value, which results
in finite error bars on these cross-over temperatures. However, a
monotonic increase of the cross-over temperatures with field is
observed, with the corresponding error bars also increasing. When
adding this cross-over line to the H-T phase diagram (see below
Fig.\ref{F09}, full triangles), it appears to converge with the
magnetic order upper boundary (solid line) around the QCP between 45
and 70 kG.

When we describe the YbPtIn low-T resistivity data as
$\rho~-~\rho_0~=~A~T^{\beta}$ for H $>$ H$^{ab}_c$, we encounter a
situation very similar to that seen in YbPt$_{0.98}$In: the
logarithmic plot in Fig.\ref{F12sqcd}a, linear at low T, is
indicative of a jump in slope for an applied field between 70 kG and
80 kG, where the slope gives the values of the exponent $\beta$. It
appears that the exponent $\beta$ has a discontinuous variation with
field (Fig.\ref{F12sqcd}b (left)) from $\beta~\simeq~1$ for H $>$
H$^{ab}_c$, to $\beta~\simeq~2$ when H $\gg$ H$^{ab}_c$. This
suggests a NFL-like regime just above the magnetically ordered
state, with a cross-over to a FL state above 100 kG, making this
compound fairly similar to the Pt-deficient one, but also to the
previously reported YbRh$_2$Si$_2$\cite{tro05} and
YbAgGe\cite{bud11}.

\begin{figure}
\begin{center}
\includegraphics[angle=0,width=90mm]{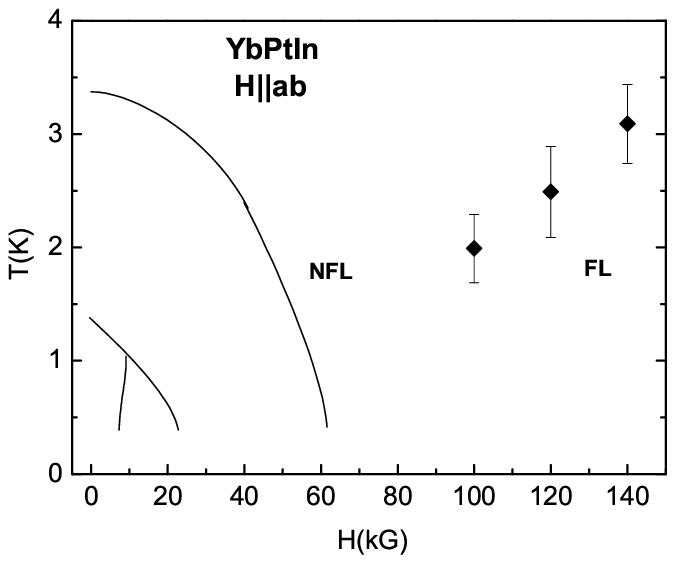}
\end{center}
\caption{Revised YbPtIn phase diagram (H $\parallel~ab$). The solid
lines correspond to the phase lines from the T - H phase diagram in
Fig.\ref{F14}, and the symbols delineate the new phase line as
determined from the $\rho(T^{\beta})$ data, using various criteria
as described in the text.}\label{F14beta}
\end{figure}

As observed before for YbPt$_{0.98}$In, due to the data scattering,
the departure of the low temperature resistivity plots from the
FL-like $\beta~\approx$ 2 behavior at the highest field allows us to
determine a cross-over temperature range, rather than a temperature
value. This results in finite error bars on the cross-over
temperatures, as seen in Fig.\ref{F12sqcd}b (right). However, an
increase of the cross-over temperatures with field is observed, both
in Fig.\ref{F12sqcd}b, and also in the revised H-T phase diagram
(Fig.\ref{F14beta}) (full symbols).

\subsection{Hall resistivity measurements in
Y\lowercase{b}P\lowercase{t}$_{0.98}$I\lowercase{n}: H$\parallel$ab}

Based on the low temperature thermodynamic and transport
measurements, YbPtIn and YbPt$_{0.98}$In can be regarded as Yb-based
heavy fermion compounds with long range, possibly reduced moment
ordering that can be driven through a field-induced quantum critical
point, similar to the previously studied Yb-based HF systems,
YbRh$_2$Si$_2$\cite{pas09} and YbAgGe\cite{bud12}. In the latter two
compounds, the Hall effect served as an additional tool for
characterizing the QCP and its effects on the finite-temperature
properties of these materials.

\begin{figure}
\begin{center}
\includegraphics[angle=0,width=90mm]{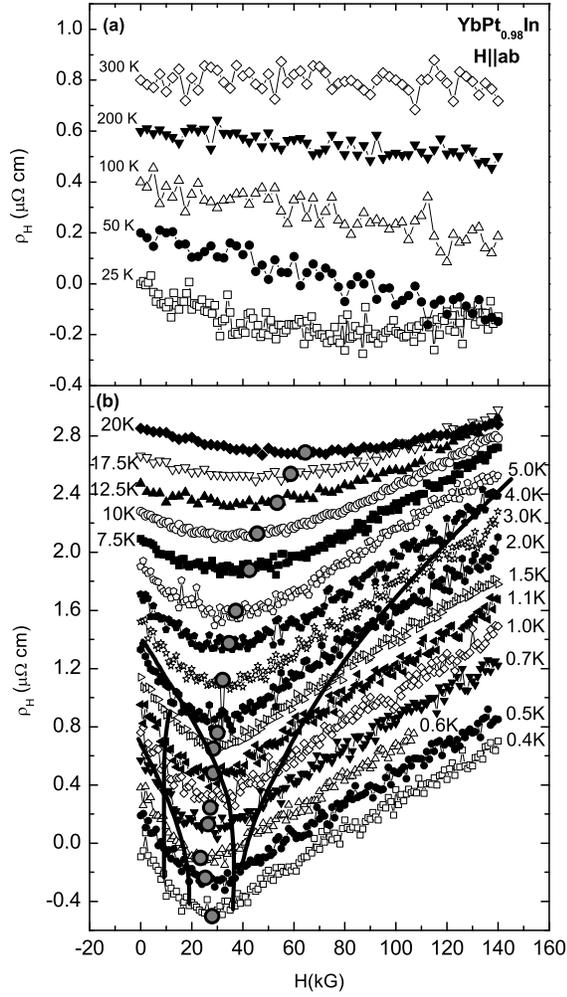}
\end{center}
\caption{Field-dependent Hall resistivity for YbPt$_{0.98}$In (H
$\parallel~ab$) for (a) T = 25 - 300 K and (b) T = 0.4 - 20 K
(except for the T = 0.4 K and 25 K, all curves are shifted up by
multiples of 0.2 $\mu \Omega$ for clarity). The lines represent the
phase transitions from the phase diagram in Fig.\ref{F06}; the large
diamonds indicate the $\rho_H$ minima.}\label{F07}
\end{figure}

In order to further explore the field-induced QCP our Yb-based
compounds, field-dependent Hall resistivity measurements were
performed for temperatures up to 300 K. However, the data on the
stoichiometric compound was noisy, given the limited (small) crystal
size, and meaningful Hall resistivity data could only be collected
for the solution-grown compound for H $\parallel~ab$. These
measurements are shown in Fig.\ref{F07}.

\begin{figure}
\begin{center}
\includegraphics[angle=0,width=90mm]{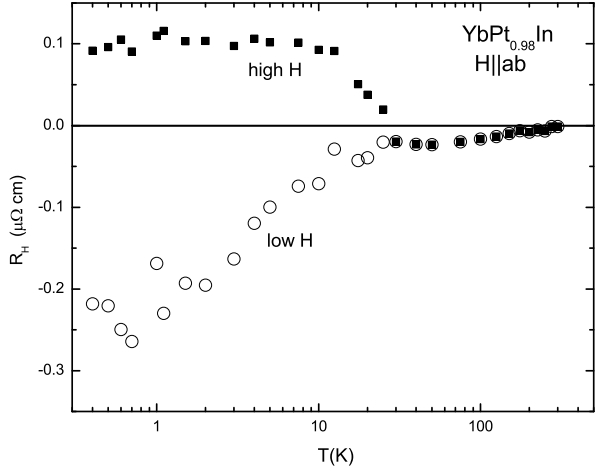}
\end{center}
\caption{Temperature-dependent Hall coefficient of YbPt$_{0.98}$In
(H $\parallel~ab$), with low-field (open symbols) and high-field
(full symbols) points determined as described in the
text.}\label{F10}
\end{figure}

\begin{figure*}
\includegraphics[angle=0,width=160mm]{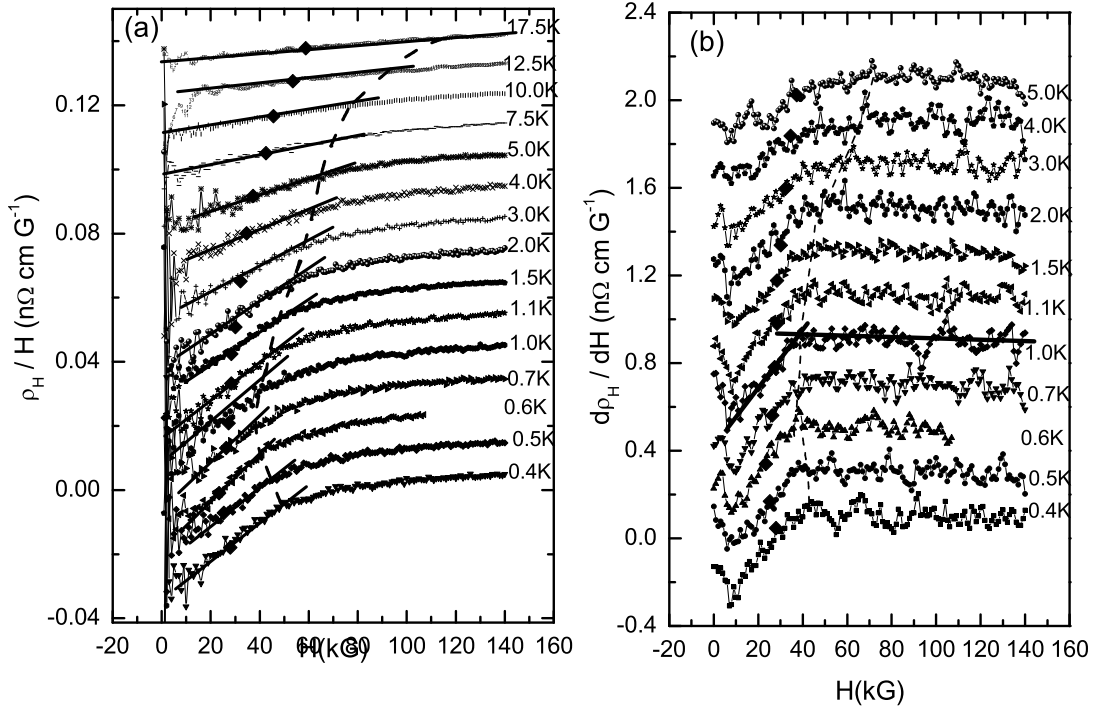}
\caption{Alternative definitions of the Hall coefficient of
YbPt$_{0.98}$In (H $\parallel~ab$): (a) R$_H$ = $\rho_H~/~H$ and (b)
R$_H$ = $d\rho_H~/~dH$, measured at various temperatures. Except for
the T = 0.4 K ones, all other curves are shifted for clarity, by
multiples of (a) 0.01 n$\Omega$ cm $/$G and (b) 0.2 n$\Omega$ cm
$/$G. The large dots are the $\rho_H$ minima shown in Fig.\ref{F07},
and the dotted line represents the cross-over line, determined by
the maximum-H on the low field linear fits (see text).}\label{F08}
\end{figure*}

We determined the Hall coefficient R$_H$ for YbPt$_{0.98}$In as
$\rho_H(H)/H$, in the low-field (H $<~30$ kG) and high-field (H
$>~60$ kG) regimes, shown in Fig.\ref{F07} for various T. (This
alternative definition for the R$_H(T)$ was preferred to
$d\rho_H~/~dH$ due to the scattering of the data at low fields as
seen later in Fig.\ref{F08}b). The results are shown in
Fig.\ref{F10} on a semi-log scale. The high-H points (squares) show
the expected leveling off of R$_H(T)$ as $T~\longrightarrow~0$,
whereas the low-H data appear to have a stronger temperature
dependence. At temperatures higher than 25 K, the two data sets
merge (as the $\rho_H(H)$ data become roughly linear for the whole
field range [Fig.\ref{F07}a]). These data indicate that the
field-dependent Hall resistivity will be non-trivial and potentially
of interest at low temperatures.

In the field-dependent Hall resistivity measurements shown in
Fig.\ref{F07}, two possible regimes can be distinguished: a high-T
regime (T $>~25$ K), where a monotonic decrease with field can be
observed, despite the scattering of the data (Fig.\ref{F07}a), and a
low-T regime (T $\leq~25$ K), for which a minimum in the $\rho_H($H)
data appears and sharpens as the temperature decreases
(Fig.\ref{F07}b). For 5 K $\leq$ T $\leq$ 25 K, the Hall resistivity
curves show a fairly broad minimum (Fig.\ref{F07}b), marked by the
large gray dots, which moves down in field with decreasing
temperature. Below 5 K, this minimum is more and more pronounced,
and is almost unaffected by the change in temperature.

Coleman \textit{et al.}\cite{col20} have indicated that R$_H$(C)
data (where C is a control parameter, \textit{i.e.,} H in our case)
can be used to distinguish between two possible QCP scenarios:
diffraction off of a critical spin density wave SDW (manifesting as
a change in the slope of R$_H$(C) at C$_{crit}$) or a breakdown of
the composite nature of the heavy electron (signaled by the
divergence of the slope of R$_H$(C) at C$_{crit}$). In our case, it
is not clear what definition of the Hall coefficient should be used
for comparison with the theory (\textit{i.e.}, either
R$_H(C)~=~\rho_H / H$ or R$_H(C)~=~d\rho_H~/~dH$) since the magnetic
field H is itself the control parameter C. Therefore in
Fig.\ref{F08} we are presenting the R$_H$(H) curves determined using
both of the aforementioned definitions. When R$_H$(H)$~=~\rho_H~/~H$
(Fig.\ref{F08}a), the Hall coefficient is linear up to a field value
which varies non-monotonically with T, as indicated by the dashed
line. This line is shifted to higher field values with respect to
the Hall resistivity line (Fig.\ref{F08}a), and is marked here by
the large diamonds. When using the R$_H$(H)$~=~d\rho_H~/~dH$
definition for the Hall coefficient (Fig.\ref{F08}b), the maximum-H
points on the low-field linear fits could be used as the criterium
for defining the Hall line (indicated by the dotted line).

Regardless of what criterium is being used, the Hall measurements
define a new phase line, distinct from any of the ones inferred from
either the C$_P$(T,H) or the $\rho$(H,T) data. Fig.\ref{HallHT}
shows the $T~-~H$ phase diagram with the phase lines determined from
the $\rho$(H,T) (H $\leq~H^{ab}_c$) or the C$_P$(T,H) data (lines),
together with the cross-over line and the Hall line inferred from
the various definitions (symbols). The line resulting from the Hall
resistivity $\rho_H$ data (diamonds) seems to persist even below
T$_{ord}$, down to our lowest T. However, the other two criteria
used for the definition of the Hall line (R$_H(H)~=~\rho_H~/~H$ or
H$_{max}$ for the linear fit of R$_H(H)~=~d\rho_H~/~dH$) delineate
lines, which more closely resembles the Hall line observed for
either YbRh$_2$Si$_2$\cite{pas09} and YbAgGe\cite{bud12} as it
appears to converge with the "coherence" line (\textit{i.e.}, the
line defining the cross-over between the non-Fermi liquid and
Fermi-liquid like regimes), and the high-T magnetic ordering phase
boundary, at the QCP, around H$^{ab}_c$. It should be noted, though
that, unlike YbAgGe, for which the Hall line is clearly not
associated with the saturation of the Yb moments\cite{tak05}, for
YbPt$_{0.98}$In the Hall line appears to be close to the saturation
fields, at least at the lowest temperatures.

\subsection{Summary}

\begin{figure}
\begin{center}
\includegraphics[angle=0,width=90mm]{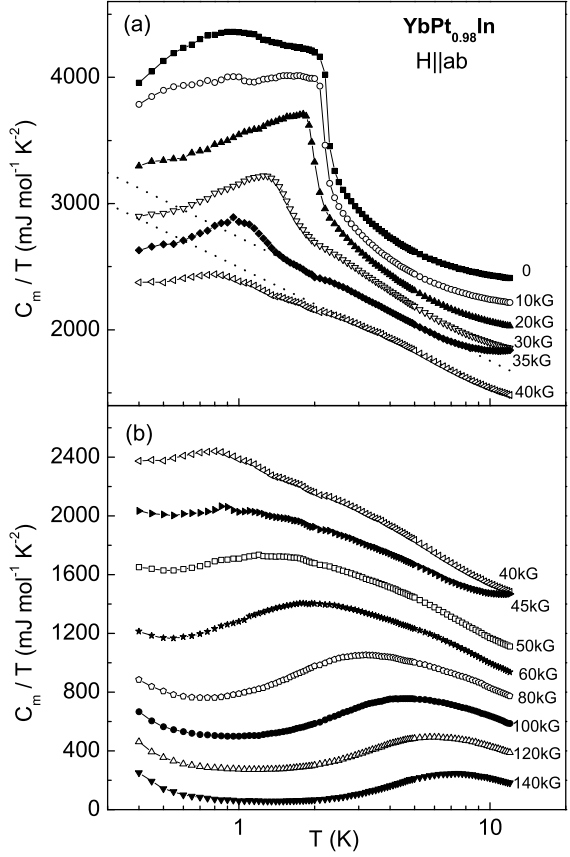}
\end{center}
\caption{Semi-log plot of C$_m/T$ \textit{vs.} T for (a) H = 0 - 40
kG and (b) H = 40 - 140 kG. All curves (except for the H = 140 kG
one) are shifted up by multiples of 200 mJ$/$mol K$^2$. The dotted
lines (for H = 35 and 40 kG) are guides to the eye for the linear
regions on the C$_m/T$ curves for H possibly just above and below
H$^{ab}_c$ (see text).}\label{F03b}
\end{figure}

\begin{figure}
\begin{center}
\includegraphics[angle=0,width=90mm]{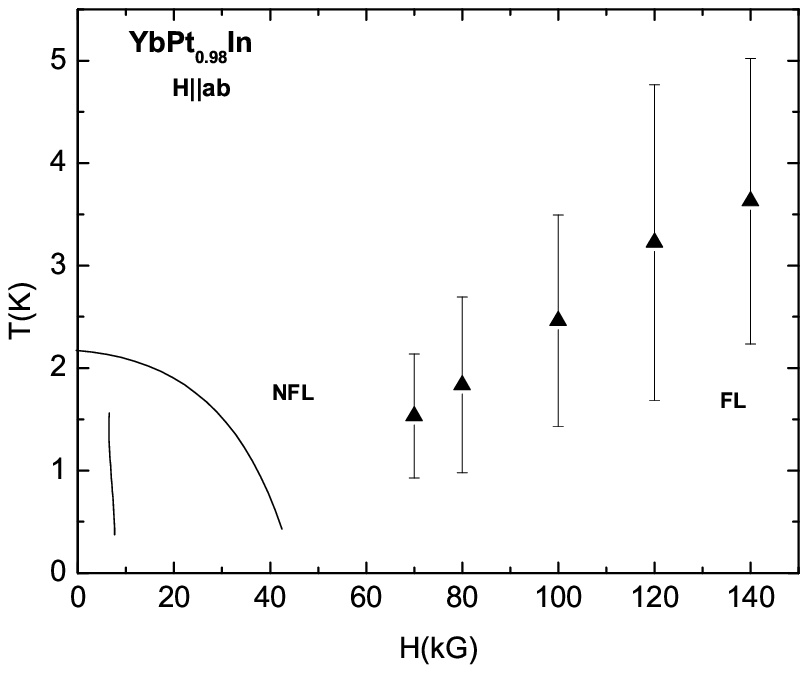}
\end{center}
\caption{Revised YbPt$_{0.98}$In phase diagram (H $\parallel~ab$).
The solid lines correspond to the phase lines from the T - H phase
diagram in Fig.\ref{F06}, and the symbols delineate the new phase
lines as determined from the $\rho(T^{\beta})$ data, using various
criteria as described in the text.}\label{F09}
\end{figure}

\begin{figure}[htb]
\begin{center}
\includegraphics[angle=0,width=90mm]{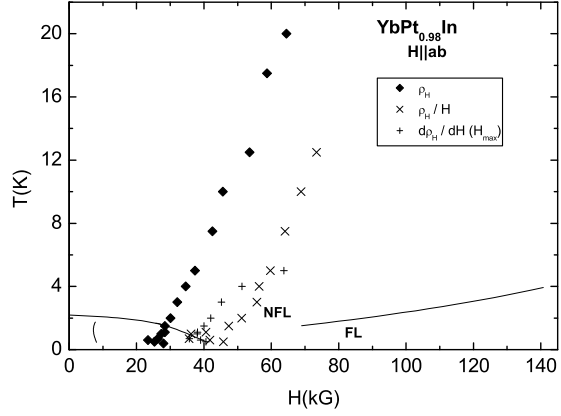}
\end{center}
\caption{Revised YbPt$_{0.98}$In phase diagram (H $\parallel~ab$).
The solid lines now represent the phase lines from the T - H phase
diagram in Figs.\ref{F06} and \ref{F09}, and the symbols delineate
the new phase line as determined from the Hall resistivity data,
using various criteria as described in the text.}\label{HallHT}
\end{figure}

Based on the presented data, we can conclude that, in both
YbPt$_{0.98}$In and YbPtIn, when magnetic field is increased above
$H^{ab}_c$, the system first enters a NFL state, followed, at higher
fields, by a FL-like state: the specific heat has a logarithmic
divergence (Figs.\ref{F03b} and \ref{F11a}) and the resistivity
shows a $\Delta \rho~\sim~A~T^\beta$ functional dependence
(Fig.\ref{F04sqcd}a,b and \ref{F12sqcd}a,b), with $\beta$ having a
discontinuous change at intermediate field values, from
$\beta~\simeq~1$ to $\beta~\simeq$ 2. Moreover, from Hall
resistivity measurements on the Pt-deficient compound, we were able
to determine another totally distinct line in the T - H phase
diagram (Fig.\ref{F09}), convergent, towards the lowest temperature,
with the cross-over line and the high-T magnetic order line. All
these observations lead us to believe that a QCP exists in both
compounds, around $H^{ab}_c~\approx~35~-~55$ kG for YbPt$_{0.98}$In,
and 60 kG for YbPtIn.

To further study the nature of the field-induced QCP in the two
systems, we analyze the field dependence of the electronic specific
heat coefficient $\gamma$. Also, in order to extend the comparison
of the disordered state in our systems and in previously studied
compounds (YbRh$_2$Si$_2$\cite{pas09} and YbAgGe\cite{bud12}), we
analyze the field-dependence of the coefficients A in the
$\Delta\rho~\propto~A T^{\beta}$ ($\beta$ = 2) resistivity, even
though in both YbPtIn and YbPt$_{0.98}$In the resistivity was not
exactly quadratic for H $\gg$ H$^{ab}_c$, but $\beta~\approx$ 2.

\begin{figure}
\begin{center}
\includegraphics[angle=0,width=90mm]{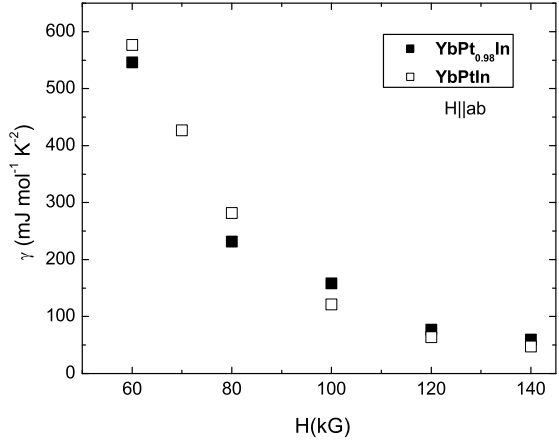}
\end{center}
\caption{The H $\parallel~ab$ field-dependent electronic specific
heat coefficient $\gamma$ of YbPtIn (open symbols) and
YbPt$_{0.98}$In (full symbols).}\label{F03a}
\end{figure}

The field-dependent electronic specific heat coefficient $\gamma$
could be estimated at low T, outside the ordered state, and these
values are shown in Fig.\ref{F03a}. For both compounds, $\gamma(H)$
was taken as the corresponding C$_m~/T$ value at T $\approx~1.3$ K,
such as to avoid the low-T upturn in the highest field data. A
drastic decrease of $\gamma$ (almost an order of magnitude) can be
observed in Fig.\ref{F03a} for both compounds, for fields above
their respective $H^{ab}_c$.

\begin{figure}
\begin{center}
\includegraphics[angle=0,width=90mm]{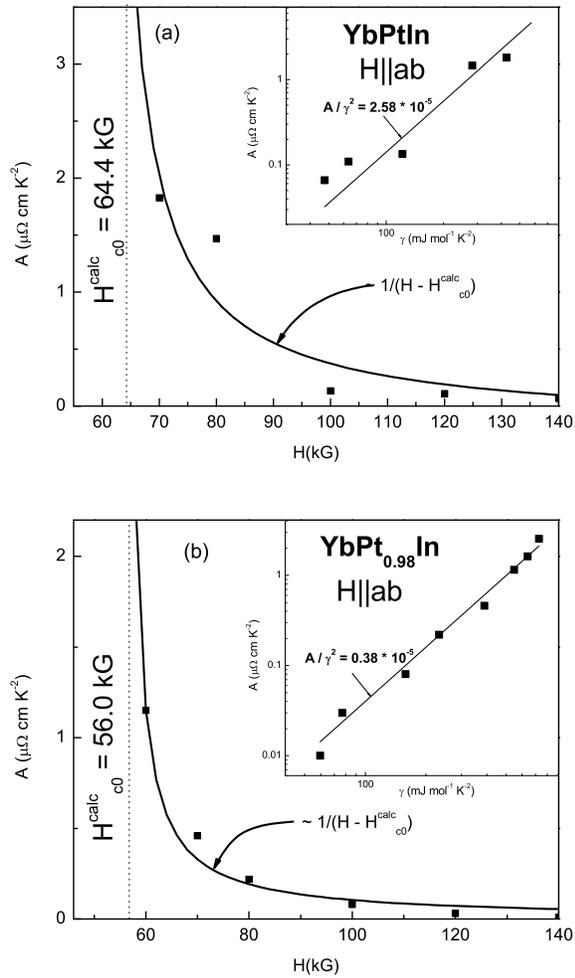}
\end{center}
\caption{The H $\parallel~ab$ T$^2$-resistivity coefficient A as a
function of field, for (a) YbPtIn and (b) YbPt$_{0.98}$In (symbols);
the solid line represents a 1$/~(H~-~H^{calc}_{c0}$) fit, from which
the expected critical field values $H^{calc}_{c0}$ were estimated
(see text). Inset: log plot of A vs. $\gamma$ (symbols), with a
linear fit (solid line) used to estimate the A $/~\gamma^2$
ratio.}\label{F23}
\end{figure}

The field dependence of the coefficient A of the quadratic
resistivity is shown in Fig.\ref{F23}a,b for YbPtIn and
YbPt$_{0.98}$In respectively. A $1~/~(H~-~H^{calc}_{c0})$ divergence
can be observed for both compounds, with $H^{calc}_{c0}$ estimated
from the fit as 64.4 kG for YbPtIn, and 56.0 kG for YbPt$_{0.98}$In.
This critical field value for the former system is consistent with
the $H~\parallel~ab$ specific heat and resistivity data
(Figs.\ref{F11}-\ref{F14}), which suggested that $H^{ab}_c~\approx$
60 kG. In the case of YbPt$_{0.98}$In, where NFL behavior was
observed between 35 kG and 55 kG, the critical field value can only
be determined to a range within the above two field values; the
$H^{calc}_{c0}~\approx~56$ kG determined from the A \textit{vs.}
$1~/~(H~-~H^{calc}_{c0})$ fit (Fig.\ref{F23}b) is close to the above
field range, thus consistent with the specific heat and resistivity
data (Figs.\ref{F03}-\ref{F06},\ref{F04sqcd}-\ref{F09}).

The proportionality $A~\sim~\gamma^2$ between the resistance
coefficient A and the electronic specific heat coefficient $\gamma$
is emphasized by the logarithmic plots in the insets in
Fig.\ref{F23}. In the case of the YbPtIn compound (Fig.\ref{F23}a,
inset), the solid line represents the Kadowaki-Woods ratio
$A~/~\gamma^2~\approx~2.6~*~10^{-5}~\mu \Omega$ cm $/$ (mJ / mol
K)$^2$. Such a value is close to that observed for many heavy
fermion systems\cite{kad21}, but slightly higher than the
$A~/~\gamma^2$ ratio reported for YbRh$_2$Si$_2$\cite{geg06}; this,
in turn, was larger than the values observed for most Yb-based
intermetallic compounds\cite{tsu22}. The corresponding value was
larger still in YbAgGe\cite{bud11}, and it would appear that this is
a common feature of Yb-based materials with field-induced NFL-like
behavior.

When we turn to the Pt-deficient compound (Fig.\ref{F23}b, inset), a
smaller $A~/~\gamma^2$ ratio is indicated by the solid line:
$A~/~\gamma^2~\approx~0.4~*~10^{-5}~\mu \Omega$ cm $/$ (mJ / mol
K)$^2$. This value is again one order of magnitude larger than that
expected for many Yb-based compounds\cite{tsu22}, and several times
smaller than the ratio observed for the stoichiometric YbPtIn
compound (Fig.\ref{F23}a). In light of observations of suppression
of the A coefficient due to site disorder\cite{tsu22,law23}, we may
have to consider the $A~/~\gamma^2$ ratio for YbPt$_{0.98}$In as a
reduced value from the enhanced ratio observed for the
stoichiometric compound.

\section{conclusions}

The detailed field- and temperature-dependent measurements presented
here allowed us to confirm that YbPtIn is a heavy fermion compound,
as has been presented by Trovarelli et al.\cite{tro09}. In addition,
we showed that a field-induced QCP exists in this material, at least
for $H~\parallel~ab$, with $H^{ab}_c~\approx~60$ kG. In addition to
the magnetic field used as a control parameter, we showed that a
small Pt-deficiency introduced in this system had effects consistent
with positive pressure applied to Yb-based heavy fermion compounds.
Thus, in the YbPt$_{0.98}$In compound we also see a suppression of
the magnetic ordering by applied magnetic field; in addition, the
small disorder also suppresses the ordered state (both T$_{ord}$ and
H$_c$ have smaller values in this compound than the similar ones in
YbPtIn).

NFL regions can be observed in both our compounds for H $>$
H$^{ab}_c$, characterized by linear $\rho(T^\beta)$ ($\beta~\simeq$
1) and logarithmic divergency of the specific heat. For H $\gg$
H$^{ab}_c$, the exponent $\beta$ approaches 2, suggesting a
cross-over to a low-T FL state at these field values, similar to
YbRh$_2$Si$_2$\cite{geg06} and YbAgGe\cite{bud11}.

As the critical field required to suppress the magnetic order in
YbPt$_{0.98}$In and YbPtIn in the c direction appeared to exceed our
maximum field, experiments to higher fields would be desirable in
order to extend the comparison with YbRh$_2$Si$_2$ and YbAgGe to the
effects of anisotropy on the field-induced QCP.

\section{acknowledgments}

Ames Laboratory is operated for the U.S. Department of Energy by
Iowa State University under Contract No. W-7405-Eng.-82. This work
was supported by the Director for Energy Research, Office of Basic
Energy Sciences.

\clearpage

\clearpage

\clearpage

\clearpage

\clearpage

\clearpage

\end{document}